\documentstyle[epsf,psfig,epsfig,12pt]{article}  
\renewcommand{\thefootnote}{\fnsymbol{footnote}}
\begin{document}
\newcommand{\be}{\begin{eqnarray}}
\newcommand{\dlq}{\lq\lq}
\newcommand{\ee}{\end{eqnarray}}
\newcommand{\ben}{\begin{eqnarray*}}
\newcommand{\een}{\end{eqnarray*}}
\newcommand{\beq}{\begin{equation}}
\newcommand{\eeq}{\end{equation}}
\renewcommand{\baselinestretch}{1.0}
\newcommand{\as}{\alpha_s}
\def\eq#1{{Eq.~(\ref{#1})}}
%%%%%%%%%%%%%%%%%%%%%%%%%%%%%%%%%%%%%%%%
% ABBREVIATED JOURNAL NAMES
\def\ap#1#2#3{     {\it Ann. Phys. (NY) }{\bf #1} (19#2) #3}
\def\arnps#1#2#3{  {\it Ann. Rev. Nucl. Part. Sci. }{\bf #1} (19#2) #3}
\def\npb#1#2#3{    {\it Nucl. Phys. }{\bf B#1} (19#2) #3}
\def\plb#1#2#3{    {\it Phys. Lett. }{\bf B#1} (19#2) #3}
\def\prd#1#2#3{    {\it Phys. Rev. }{\bf D#1} (19#2) #3}
\def\prep#1#2#3{   {\it Phys. Rep. }{\bf #1} (19#2) #3}
\def\prl#1#2#3{    {\it Phys. Rev. Lett. }{\bf #1} (19#2) #3}
\def\ptp#1#2#3{    {\it Prog. Theor. Phys. }{\bf #1} (19#2) #3}
\def\rmp#1#2#3{    {\it Rev. Mod. Phys. }{\bf #1} (19#2) #3}
\def\zpc#1#2#3{    {\it Z. Phys. }{\bf C#1} (19#2) #3}
\def\mpla#1#2#3{   {\it Mod. Phys. Lett. }{\bf A#1} (19#2) #3}
\def\nc#1#2#3{     {\it Nuovo Cim. }{\bf #1} (19#2) #3}
\def\yf#1#2#3{     {\it Yad. Fiz. }{\bf #1} (19#2) #3}
\def\sjnp#1#2#3{   {\it Sov. J. Nucl. Phys. }{\bf #1} (19#2) #3}
\def\jetp#1#2#3{   {\it Sov. Phys. }{JETP }{\bf #1} (19#2) #3}
\def\jetpl#1#2#3{  {\it JETP Lett. }{\bf #1} (19#2) #3}
\def\epj#1#2#3{    {\it Eur. Phys. J. }{\bf C#1} (19#2) #3}
\def\ijmpa#1#2#3{  {\it Int. J. of Mod. Phys.}{\bf A#1} (19#2) #3}
%%%%%%%%% notice the parentheses is only on one side
\def\ppsjnp#1#2#3{ {\it (Sov. J. Nucl. Phys. }{\bf #1} (19#2) #3}
\def\ppjetp#1#2#3{ {\it (Sov. Phys. JETP }{\bf #1} (19#2) #3}
\def\ppjetpl#1#2#3{{\it (JETP Lett. }{\bf #1} (19#2) #3} 
\def\zetf#1#2#3{   {\it Zh. ETF }{\bf #1}(19#2) #3}
\def\cmp#1#2#3{    {\it Comm. Math. Phys. }{\bf #1} (19#2) #3}
\def\cpc#1#2#3{    {\it Comp. Phys. Commun. }{\bf #1} (19#2) #3}
\def\dis#1#2{      {\it Dissertation, }{\sf #1 } 19#2}
\def\dip#1#2#3{    {\it Diplomarbeit, }{\sf #1 #2} 19#3 }   
\def\ib#1#2#3{     {\it ibid. }{\bf #1} (19#2) #3}
\def\jpg#1#2#3{        {\it J. Phys}. {\bf G#1}#2#3}

\begin{flushright}
TAUP-2665-2001\\
\today

\end{flushright}

\vspace*{1cm} 
\setcounter{footnote}{1}
\begin{center}
{\Large\bf High density QCD at THERA }
\\[1cm]

 E. \ Gotsman$^{a}$, E. \ Levin$^{a}$, \   M. \ Lublinsky$^{b}$, \ U. \
Maor$^{a}$, \ 
  E. \ Naftali$^{a}$ and  K. \ Tuchin$^{a}$\\  
~~ \\
~~\\
{\it ${}^{a}$ HEP Department, School of Physics and Astronomy } \\ 
{\it Tel Aviv University, Tel Aviv 69978, Israel } \\ 
 ~~ \\
{\it ${}^{b}$Department of  Physics, Technion,}\\
{\it Haifa, 32000, Israel}\\
~~\\
~~\\
 
\end{center}
\begin{abstract} 

These notes are a summary of our predictions for the new THERA project,
related to deep inelastic scattering in the region of ultra low $x$ (
$x \rightarrow 10^{-7}$). We collect here  predictions that
satisfy two criteria (i) they do not depend on specific features of
the model that we have to use to estimate a possible effect; and (ii) 
they do not contradict the  HERA data.

\end{abstract}
\renewcommand{\thefootnote}{\arabic{footnote}}
\setcounter{footnote}{0}

\section{Introduction: Our hopes and main goals at THERA}
\subsection{Three domains of QCD at low $\mathbf{x}$}

Deep inelastic scattering is a unique experiment which allows us to take 
`snapshots' of the constituents inside a hadron at different moments of  time with 
different resolutions. These `snapshots' provide  the  possibility of finding
the 
degrees of freedom (DOF) that are responsible for the  interaction in QCD
and
generalize  the theoretical  approach from the well defined domain of
perturbative QCD
to  the  unknown non-perturbative ( confinement ) region, where the
appropriate theoretical methods are still to be determined.
 DIS
allows one  to see the constituents of  size $\approx  1 /Q$, where $Q$
is the photon virtuality, at the time  $t \approx 1/m x$,  where
$x$ is the Bjorken variable related to the energy ($W$) of the process
($x = Q^2/W^2$ at low $x$ ).

\begin{figure}[h]
\begin{minipage}{10.0cm}
\begin{center}
\epsfysize=8.4cm
\leavevmode
\hbox{ \epsffile{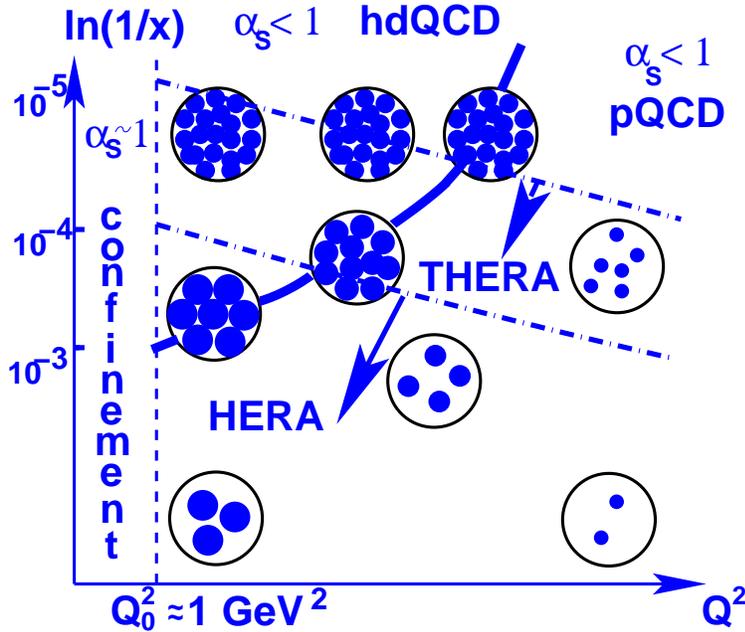}}
\end{center}
\end{minipage}
\begin{minipage}{5.0cm}
\caption{\it `Snapshots' of the constituents of different size ($r \approx
1/Q$) at different values of $x$. Large  circles indicate  the hadron and
small  ones  the constituents.}
 \label{sat}
 \end{minipage}
\end{figure}

HERA data as well as theoretical studies suggest that hadrons have
qualitatively diverse  structure in the  three different  domains ( see
Figs.~\ref{phase} and \ref{sat} ):
\begin{figure}[h]
\begin{minipage}{12.0 cm}
\begin{center}
\epsfysize=8.4cm
\leavevmode
\hbox{ \epsffile{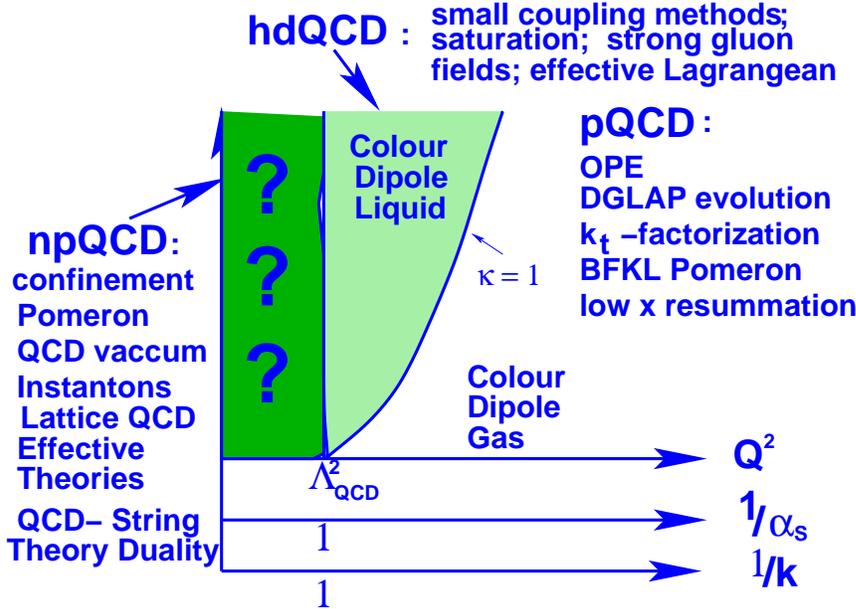}}
\end{center}
\end{minipage}
\begin{minipage}{4.0cm}
\caption{\it Phase diagram of DIS. $\kappa$ is the packing factor which is
the number of constituents multiplied by its typical area and divided by
the area of a hadron.}
\label{phase}
\end{minipage}   
\end{figure}

\begin{figure}[h]
\begin{minipage}{11.0cm}
\begin{center}
\epsfysize=8.4cm
\leavevmode 
\hbox{ \epsffile{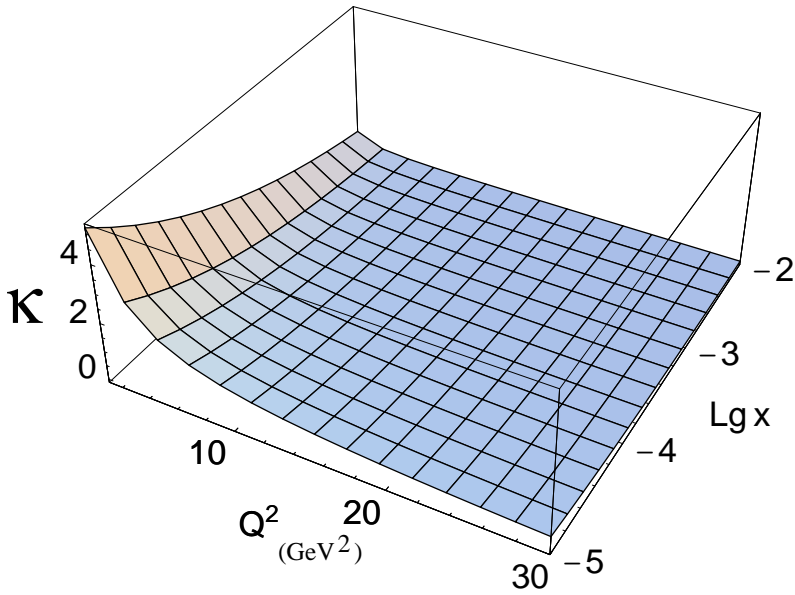}}
\end{center}
\end{minipage}
\begin{minipage}{5.0cm}
\caption{\it  The parton packing factor $\kappa$ as function of $Q^2$ and 
Bjorken $x$ in the GRV'98 parameterization of the solution to the DGLAP
evolution equation. The  GRV'98 parameterization describes  all 
available data from HERA.}
 \label{kappa}
 \end{minipage}
\end{figure}

\begin{enumerate}
{\bf \item\,\,\,Perturbative QCD domain} where the constituents are  of small
size and 
are 
distributed in a hadron with rather low density ( packing factor of these
constituents $\kappa$ is small (  $\kappa < 1 $ , see Fig.~\ref{kappa} ));

{\bf \item\,\,\,High parton density QCD domain} in which the  constituents
are still small  and we can use  weak coupling methods, but
their density is so large that their packing
factor $\kappa > 1 $, and so we cannot treat this system of partons using the
established  pQCD methods;

{\bf \item\,\,\, Non perturbative QCD domain}  in which the  QCD coupling is
large, the confinement of quarks and gluons occurs, and new theoretical
methods must  be developed to  explore this region.

\end{enumerate}

Fig.~\ref{sat} illustrates  these `snapshots' of the  constituents at
different moments of time (
different values of $x$ ). One can see three domains with different
distributions of the constituents in the transverse plane. It should be
 stressed that the distributions do not depend on the reference frame,
unlike time which differs in  different  reference  frames.
   
Each of these domains has its own   theoretical problems that can be 
clarified by THERA experiments. The key problems are shown in
Fig. ~\ref{phase}. 

\subsection{Brief summary of HERA data}

\begin{figure}[h]
\begin{minipage}{10.0cm}
\begin{center}
\epsfxsize=8.0cm
\epsfysize=8.0cm
\leavevmode
\hbox{\epsffile{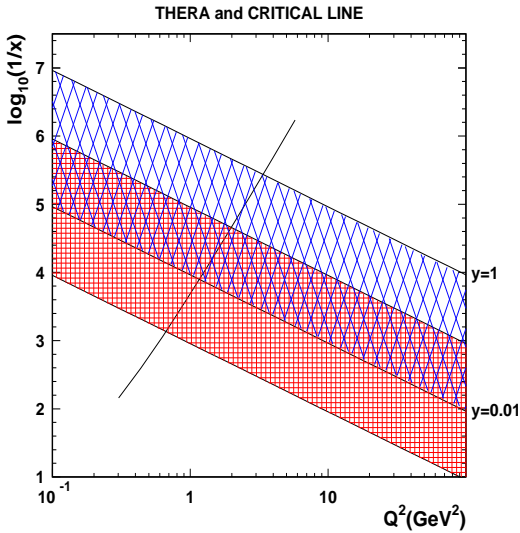}}
\end{center}
\end{minipage}
\begin{minipage}{5.0cm}
\caption{\it The estimates for the saturation scale for HERA and THERA
kinematic region.}
 \label{satscale}
 \end{minipage}
\end{figure}

Brief resume of HERA data: these data can be  described by 
models including   parton  saturation,  but they can also  be described without
assuming  saturation.  However, it  turns out that all predictions of  asymptotic
 hdQCD have already  been seen  in HERA data. This fact is so impressive and convincing that we, personally,
think that HERA has reached a new regime of high density QCD \cite{AMIRIM}.
 However,  the situation
 is still  non-conclusive as is  illustrated by Fig.~\ref{satscale} which shows the value of the
saturation scale in HERA and THERA kinematic region. One can see that
$Q_s(x) \leq 1 \,GeV$ for HERA. This low  $Q_s(x)$ indicates that HERA
data can be described by other approaches without  saturation, for
example, by some models that include a smooth matching between
``soft" and
``hard"  interactions. 
 However, at THERA $Q_s(x)$ is larger and the hdQCD
interpretation of the data will be cleaner.  We illustrate this point
with
two figures that follow.

\subsection{Main idea} 

As we have discussed,  we face two challenging problems
in the region of low $x$ and low $Q^2$  which is now being investigated at 
HERA 
:
\begin{enumerate}
\item\,\,\, The matching of ``hard" processes, which can be successfully
described using perturbative QCD (pQCD),  and ``soft" processes, which 
should
be described using non-perturbative QCD (npQCD);        

\item\,\,\, Theoretical description of high density QCD
(hdQCD). In this kinematic region we expect that the
typical distances will be small,  but the parton density will be so large
that a new non perturbative approach needs to  be developed for
dealing with
this system.
\end{enumerate}
 The main  physical idea, on which our approach is
based  is  \cite{SAT}:
                               
{ \large \em
The above two problems are correlated
and the system of partons always passes through the stage of hdQCD
( at shorter distances ) before it proceeds to 
non-perturbative QCD and which, in  practice, we describe using  Reggeon phenomenology.}     

\subsection{Status of theory }

Parton  saturation as well as other collective phenomena 
typical of  the high parton density system, is not an additional
postulate of QCD,
but follows from the QCD evolution equations in  the 
kinematic region  associated with high parton density. Therefore, it is very important to have a clear
understanding what can be proven theoretically. 
 
In DIS at low $x$, one can find a system of  high density partons,
which is a non-perturbative system due to high density of partons,
although  the
running QCD coupling constant is still small ( $\alpha_S(r_{\perp}) \ll 1
$ ). Such a unique system can be treated theoretically \cite{SAT}. It
should be stressed that the theory of hdQCD is now in  very good shape.   

Two approaches have been developed for hdQCD. The first one
\cite{PTHEORY} is based on pQCD ( see GLR and Mueller and Qiu papers in
Ref. \cite{SAT} ) and on dipole degrees of freedom \cite{MU94}. This
approach gives a natural description of the parton cascade in the
kinematic region for  $\kappa \leq 1$  and up to the
transition region with $\kappa \approx 1 $ ( see Fig. ~\ref{sat}).

The second method \cite{ELTHEORY} uses the effective Lagrangian suggested by McLerran and
Venugopalan \cite{SAT}, this is a natural framework to describe data
in the deep
saturation region,  where $\kappa \gg 1$ (see Fig. ~\ref{sat}).
As a result of intensive work using  these two approaches the
non-linear evolution equation which has the following form  has been derived \cite{EQ}                          
                         
\be \label{GLRINT}
\frac{d a^{el}({\mathbf{x_{01}}},b_t,y)}{d y}\,\,\,&=&\,\,\,- \,\frac{2
\,C_F\,\as}{\pi} \,\ln\left(
\frac{{\mathbf{x^2_{01}}}}{\rho^2}\right)\,\,
a^{el}({\mathbf{x}},b_t,y)\,\,\,+
\,\,\,\frac{C_F\,\as}{\pi}\,\,
\int_{\rho} \,\,d^2 {\mathbf{x_{2}}}\,
\frac{{\mathbf{x^2_{01}}}}{{\mathbf{x^2_{02}}}\,
{\mathbf{x^2_{12}}}}\, \\
 &\cdot&\,\,\,
\left(\,\,2\,a^{el}({\mathbf{x_{02}}},b_t,y)\,
-\,a^{el}({\mathbf{x_{02}}}, b_t,y)\,a^{el}({\mathbf{x_{12}}},
 b_t,y)\,\right)\,\,, \nonumber
\ee
where $a^{el}(r^2_{\perp},b_t,x)$ is the elastic  scattering amplitude for a
dipole      of size $r_{\perp}$ at energy $\propto 1/x$ and at impact
parameter
$b_t$. We assume that $b_t \,\,\leq x_{02} \,\,\mbox{and/or}
\,\,x_{12}$.

\begin{figure}[h]
\begin{minipage}{12.0 cm}
 \begin{center}
\epsfxsize=11.5cm
\epsfysize=7cm
\leavevmode
\hbox{ \epsffile{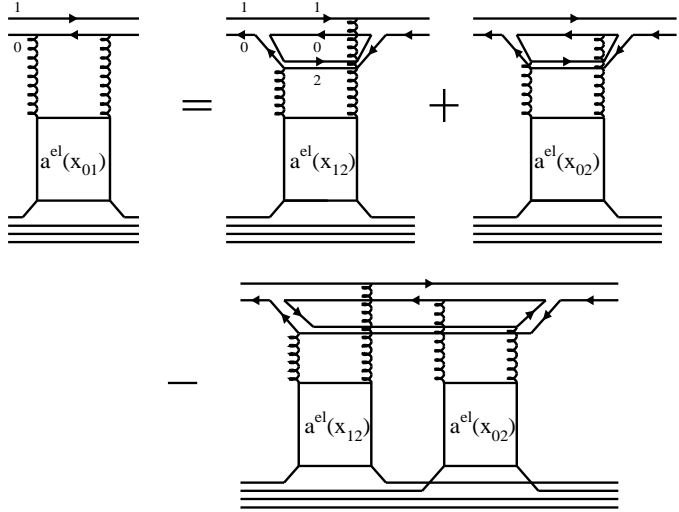}}
\end{center}
\end{minipage}
\begin{minipage}{4.0 cm} 
\caption{\it The pictorial form of the non-linear evolution equation in
the hdQCD
kinematic region.}
\label{hdqcd}
\end{minipage} 
\end{figure}

 The dipole cross section  is equal to $\sigma(r^2_{\perp},x)
=\,2 \,\int \,d^2 b_t \,a^{el}(r^2_{\perp},b_t,x)$. The pictorial form of
\eq{GLRINT} is given in Fig.~\ref{hdqcd} which shows that the physics
underlying this equation
has a  simple  meaning: the dipole of size $x_{10}$ decays
in two dipoles of  sizes $x_{12} $ and $x_{02}$. These two dipoles
interact with the target. The non-linear term which takes into account
the
Glauber corrections for such an interaction, \eq{GLRINT} is the same as
the GLR -equation  \cite{SAT} but here it is given  in the  coordinate
representation. The  coefficient in front of the non-linear term coincides in
the double log  with
the one       calculated in Ref. \cite{AGL}.
 
 We wish to stress
 that this equation which includes the Glauber rescatterings
, has definite initial conditions and has been derived by both methods
 (see Refs. \cite{EQ,LARRY}).        
The model, which we will describe in the next section, provides both the
correct initial conditions for \eq{GLRINT} and also serves as a good first
iteration. This iteration reproduces the main features of the solution,
and it is only necessary  to repeat the iteration procedure  two or three times to
obtain a correct solution for $x \leq 10^{-7}$.

\section{Our model}
\subsection{General description}
As was shown in Ref. \cite{EQ}, the correct initial condition for
\eq{GLRINT} is actually the Glauber-Mueller formula \cite{DOF1,DOF2,DOF3} for the 
rescattering of the colour dipole, namely
 
\beq \label{GM1}
\sigma_{dipole}(r_{\perp},x) \,\,=\,\,2\,\,\int\,d^2 b_t\,Im
\,a^{el}(r_{\perp},x; b_t) \,\,,
\eeq
where
\beq \label{GM2}
a^{el}(r_{\perp},x; b_t)\,\,= i \left(\,1\,\,-\,\,e^{ -
\frac{\Omega(r_{\perp},x;b_t)}{2}}\,\right)\,\,,
\eeq  

The opacity $\Omega(r_{\perp},x;b_t)$ is defined as 
\beq \label{OMEGA}
\Omega(r_{\perp},x;b_t)\,\,\,=\,\,\,\frac{\pi^2 r^2_{\perp}}{3 \pi
R^2}\,x
G^{DGLAP}(x,  \frac{4}{r^2_{\perp}};b_t)\,\,,
\eeq
where $G^{DGLAP}(x,  \frac{4}{r^2_{\perp}};b_t) = G^{DGLAP}(x,
\frac{4}{r^2_{\perp}}) \cdot S(b_t)$ and  $G^{DGLAP}(x,  \frac{4}{r^2_{\perp}})$
 is the solution of the linear DGLAP evolution equation,  and $S(b_t)$ is the
profile function for the impact parameter distribution of the  gluons in the
target. The origin of this function is   non-perturbative,  and it is normalized
in \eq{OMEGA} by the condition $S(b_t=0) = 1$.
   
For the solution of \eq{GLRINT} one  should fix the value of initial $x=x_0$
 ($y = y_0$) and use $a^{el}(r_{\perp},x= x_0; b_t)
$ as a starting iteration of \eq{GLRINT}. In our model we suggest a different
 approach, namely we use \eq{GM1} and \eq{GM2} as the  first iteration of
 the \eq{GLRINT} including their $x$-dependence.  Therefore, the result of the 
second iteration of \eq{GLRINT} can be written in the form:
\beq \label{2IT}
a^{el}_2(r_{\perp},y= \ln(1/x); b_t)\,\,=\,\,
\frac{C_F\,\as}{\pi}\,\, r^2_{\perp}
\int^y dy'\, \int_{r_{\perp}} \,\,\frac{d^2 r'_{\perp}}{r'^4_{\perp}}\,
\,\,\{\, 2\,a^{el}_1(r'_{\perp},y'; b_t)\,\,-\,\,(\,a^{el}_1(r'_{\perp},y'
; b_t)\,)^2\,\}\,\,,
\eeq
where we assumed that $r'_{\perp} \,\,\gg\,\,r_{\perp}$. This assumption corresponds to
 the Leading Log  Approximation of pertutbative QCD LLA, which has been used
in the derivation of \eq{GM1} and \eq{GM2}.
Substituting \eq{GM1} in \eq{2IT} we obtain
\beq \label{GSC}
a^{el}_2(r_{\perp},y= \ln(1/x); b_t)\,\,=\,\,r^2_{\perp}\,\int^y dy' \,\,
\int_{
r^2_{\perp}}\,\,\frac{ d\,r'^2_{\perp}}{r'^4_{\perp}} \,  \left( \, 1
\,\,-\,\,e^{ -
\frac{\Omega_G(r'_{\perp},y';b_t)}{2}}\,\right)\,,
\eeq
where  $\Omega_G(r_{\perp},y;b_t) = 2   \Omega(r_{\perp},y;b_t)$ of \eq{OMEGA}
\footnote{Actually, $ \Omega_G/\Omega\,\,=\,\,2 N^2_c/(N^2_c - 1 ) = 9/4 (N_c = 3)
 \rightarrow 2 ( N_c \gg 1 )$.}.

The physical meaning of \eq{GSC} is  transparent. The dipole of  size 
$r_{\perp}$ decays into two dipoles which interact with the target. \eq{GSC} 
describes the  rescatterings of these two dipoles. On the other hand, in pQCD this
 state is the $q \bar q G $ state. Since we assume  the size of $q \bar q $ system 
to be  much smaller than the size of the dipoles in the       $q \bar q G $
state, \eq{GSC} relates
to the passage of the gluon through the target. 
Using the relation between the gluon-target cross section\footnote{In
principle, the gluon - target cross sections can be measured using the
graviton as a colouless probe.} and the gluon
distribution
 $$ \sigma^G = 2\,\int 
\,d^2 b_t   a^{el}_2(r_{\perp},y= \ln(1/x); b_t) = \frac{4 \pi^2}{Q^2}
 \alpha_S(Q^2) \,x\,G(x,Q^2)$$ one obtains the Glauber-Mueller formula for the 
gluon distribution \cite{DOF3} 
\beq \label{G}
x G^{SC}(x, Q^2) \,\,=\,\,\frac{8}{\pi^4}\,\,\int^1_x\,\frac{d x'}{x'}
\,\,\int_{4/Q^2} \,\frac{ d^2 r'_{\perp}}{r'^4_{\perp}}\,\,\int\,\,d^2\,b_t
\,\,\left(\,1\,\,-\,\,e^{ - \frac{\Omega^G}{2}}\,\right)
\,\,.
\eeq                                                                  

\eq{GM1} and \eq{G} are the  main formulae that we use  in our estimates of the
 collective phenomena in DIS.

\subsection{ Advantages and disadvantages   of the model.}

The main advantages of our model follow directly from the way it has been constructed.
Our model reproduces the DGLAP limit for
$r^2_{\perp} \,<\, r^2_{saturation}  \approx 1/Q^2_s$, gives a good approximation to
 the solution of \eq{GLRINT} for $x \,\geq \,10^{-6}$, and it preserves the relation
 between elastic, quasi-elastic ( diffraction) scattering and multi particle
 production in DIS based on the AGK cutting rules \cite{AGK} (see Ref. \cite{AMIRIM} 
for details).

The main  problem relating to   our model is the fact that  the evolution equation 
\eq{GLRINT} has only  been proven in the leading ln(1/x) approximation 
of pQCD where  we consider $\alpha_S \ln(1/x) \approx 1$ while $  \alpha_S
\ll 1$. This approximation does not insure the accuracy of calculation
for present accessible energies. On the other had,
  our model cannot be correct at low $x$ and it  is only suitable  to describe DIS for
 $x \geq 10^{-6}$, where  the model gives  the second iteration of
\eq{GLRINT}. For smaller values of $x$ we require higher iterations.
          
\begin{figure}[hptb]
\begin{center}
\begin{tabular}{ c c}
\psfig{file=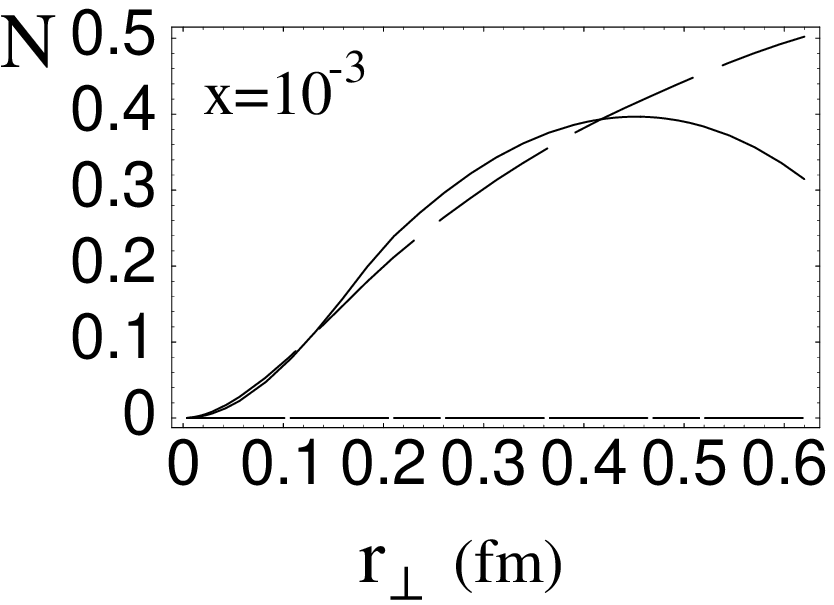,width=70mm} & \psfig{file=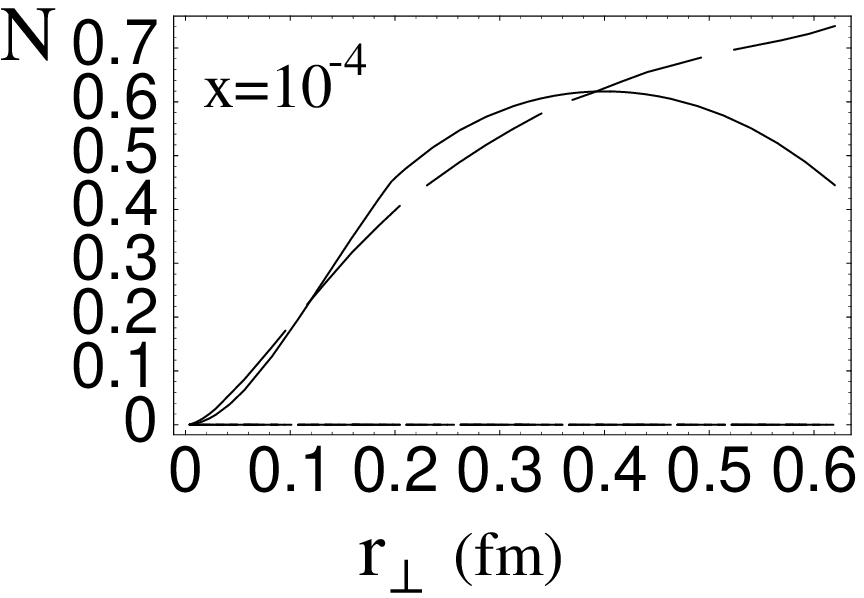,width=70mm}\\
Fig.6-a & Fig. 6-b \\
\end{tabular}
\end{center}
\caption{\it Figs.6-a - 6-b show the calulation for $N= Im\, a^{el}$ at $b_t
=0$ ($
\sigma_{dipole} = \int \,d^2 b_t \,N(x,r_{\perp};b_t) $)  in our model (
full curve ) and the solution to nonlinear equation (see \eq{GLRINT}).}
\label{neq}
\end{figure}

In Ref. \cite{LUB} we have already  shown that our model gives a good approximation of \eq{GLRINT}.
 Fig. \ref{neq} illustrates this point. Hence  we can safely use our model
 for estimates of the collective phenomena  even in THERA kinematic region.

Table 1 provides a guide for the different processes which we described in
our model for kinematical range at HERA.

\begin{table}
\centerline{\bf Table 1}
\begin{tabular}{l l l l}
Reaction & $Q^2\, ( GeV^2) $ & $x$ & References \\
\hline
$\sigma_{tot}(\gamma^* p )$ & $0 \div 65$ & $ < 0.01$ & \cite{GLMPHP} \\
   &  & & \\
$F_2(x,Q^2)$ & $1 \div 65 $ & $< 0.01 $ & \cite{AGL,GLMF2} \\
  &  & & \\
$xG(Q^2,x)$ & $1 \div 65 $ & $< 0.01 $ & \cite{AGL} \\
  &  & & \\
$d F_2/d ln Q^2$ & $ 1 \div 65 $ & $<  0.01 $ & \cite{GLMSLP,OSAKA}\\
  &  & & \\
$\sigma_{tot} ( \gamma\, \gamma^* ) $ & $0; 0 \div 20$ & $ < 0.01$ &
\cite{GLMPHPH}\\   
  &  & & \\
$\sigma^{diff}_{tot}$ & $5 \div 65$ & $ < 0.01$ & \cite{GLMINDD} \\
 &  & & \\
$\frac{\sigma^{diff}_{tot}}{\sigma_{tot}}$ & $1 \div 65$ & $ < 0.01$
&\cite{GLMRDT}\\
 &  & & \\
$\sigma( \gamma^* p \rightarrow J/\Psi + p ) $ & $0  \div 65$ & $ < 0.01$
& \cite{GLMVP,OSAKA} \\
 &  & & \\
slope $B (  \gamma^* p \rightarrow J/\Psi + p ) $ & $0  \div 65$ & $ <
0.01$ &  \cite{GLMVP} \\
 &  & & \\
slope $B (  \gamma^* p \rightarrow \rho  + p ) $ & $5  \div 65$ & $ <
0.01$   &  \cite{GLMVP}\\
\end{tabular}
\end{table}

\subsection{Phenomenological parameters of the model:}
 
Before discussing the applications at  THERA  we  list the
parameters that we use to fit the data at HERA.

\subsubsection{ $\mathbf{ R^2}$ - size of the target.}
 
The size of the target enters the impact parameter profile of the target
which we take in the Gaussian form:
\begin{equation} \label{M1}
S(b_t) \,\,=\,\,\frac{1}{\pi R^2}\,e^{ - \frac{b^2_t}{R^2}}\,\,.
\end{equation}
The HERA data for photo production of the  J/$\Psi$ - meson as well as CDF data
on double parton cross section, leads to the value of $R^2 = 5 \div
10\,\,GeV^{-2}$. 
$R^2$ is a  parameter fiied to
 describe    of the experimental data. Note, that the value of
$R^2= 8.5\,GeV^{-2}$ was taken  for all reactions that we have
described.
 
\subsubsection{ $\mathbf{ Q^2_0 = 1/r^2_{sep}}$ - separation
parameter.}
                                            
As we have discussed we can only rely on  our model for  the saturation effect (
see
\eq{GM1} - \eq{GM2} )    at rather small distances
($r_{\perp} \,<\,r^{sep}_{\perp}$ ) or , in other words, at large
virtualities of the incoming photon $Q^2 > Q^2_0$. We have commented on
the
value of  $r^{sep}_{\perp}$, but in practice we used $Q^2_0 = 0.6
\div  1
\,\,GeV^2$ and tried to estimate  how our fit depends on the value of $Q^2_0$.
Therefore, the result of our calculations should be read , as
{\it ``the shadowing  corrections from short  distances $r_{\perp} <
1/Q^2_0$ gives this or that ...."}.    
 
\subsubsection{ Solution of the DGLAP  evolution equations.}

We attempted  to use all available parameterization of the solution of the
DGLAP evolution equations\cite{CTEQ,MRS}, but we  prefer  the GRV
parameterization\cite{GRV} . The reason for this is very simple: the
theoretical
formulae, that are  the basis of our model,  were derived in double log
approximation of pQCD,  and the GRV parameterization is  the closest one to
the DLA.

\section{Predictions for THERA}
\subsection{The unitarity bound in THERA kinematic region}
We start from the prediction which, in principle, does not depend on the exact form of 
the correct evolution equation and/or on the particular model, namely, from unitarity bound 
for $F_2$ and $xG(x,Q^2)$\cite{AGLFRST}. This bound stems from a  simple formula for the DIS cross section
 \cite{DOF1,DOF2,DOF3}
\beq \label{DISXS}
\sigma(\gamma^* p ) \,\,=\,\,\int^1_0\,d z \,\int \,d^2 r_{\perp} |\Psi(z, r_{\perp}; Q^2)|^2 \,\,
\sigma_{dipole}(x_B,r^2_{\perp})\,\,,
\eeq
where $ \sigma_{dipole}(x_B,r^2_{\perp})$ is the total cross section of the
$q \bar q $ -dipole of  size
 $r_{\perp}$ with the target; $\Psi$ is the wave function of the $q \bar q $ -dipole in the virtual photon. 
This wave function is well known \cite{DOF3,WF} and for transverse polarised photon 
$|\Psi_T(z, r_{\perp}; Q^2)|^2$ is equal 
\beq \label{PSIT}
|\Psi_T(z, r_{\perp}; Q^2)|^2\,\,=\,\,\frac{\alpha^{em} N_c}{2 \pi^2}\,\times\,\sum^{N_f}_{1}
\,Z^2_f\,\,[\,z^2\,+\,(1 - z )^2\,]\,\,\tilde{Q}^2\,K^2_1(\tilde{Q}\,r_{\perp})\,\,,
\eeq
where $K_1$ is the modified Bessel function, $ \tilde{Q}^2\,=\,z(1 - z ) \,Q^2$, $N_f$ is the number of 
massless quarks and $Z_f$ is the fraction of the charge carried by the quark.

It was shown in Ref. \cite{DOF3,AGLFRST} that in the DGLAP limit the essential $r_{\perp}$ in \eq{DISXS}
 are larger than   $ 2/Q$ ( $
r_{\perp}\,>\,2/Q$) and the integral over $z$ can be taken, namely,
\beq \label{Z}
\int^1_0\,d z  |  \Psi_T(z, r_{\perp}; Q^2)|^2  \,\,\rightarrow\,\,\frac{8}{3\,Q^2\,r^4_{\perp}}\,\,.
\eeq                       
 Finally, using the relation between the total cross section and  $ F_2$  structure function 
\beq \label{F2}
 \sigma(\gamma^* p )\,\,=\,\,\frac{4\,\pi^2\,\alpha^{em}}{Q^2}\,F_2(x_B,Q^2)
\eeq
one  obtains
\beq \label{F2DIPOLE}
 F_2(x_B,Q^2)\,\,=\,\,\frac{N_c}{12 \pi^3}\,\, \sum^{N_f}_{1}\,\,Z^2_f  \,\,\int^{\infty}_{\frac{1}{Q^2}}
    \,\frac{d\,r^2_{\perp}}{ r^4_{\perp}}\,\,=\,\,\frac{N_c}{12 \pi^3}\,\, \sum^{N_f}_{1}\,\,Z^2_f
\,\,2\,\int \,d^2 b_t \,\,Im\,a^{el}_{dipole}( x_B,r_{\perp};b_t)
\eeq

Taking the derivative with respect to $\ln Q^2$ and using the weak form of the unitarity constraint ($ 
Im\,\,\,a^{el}_{dipole}( x_B,r_{\perp};b_t)\,\,\leq\,\,1$ )  we obtain

\beq \label{UN1}
\frac{\partial F_2(x,Q^2)}{\partial \ln Q^2}\,\,\leq\,\,\,  \frac{Q^2\,R^2}{3 \pi^2}\,\,.
\eeq
$R^2$ in \eq{UN1} is the region of convergence for the  integral over $b_t$ in \eq{UN1}. In principle, $R^2$ 
grows with $x$ but model estimates \cite{AGLFRST} as well as experimental data \cite{HERAREV}
 show only mild $x$  dependence.

\begin{figure}
\begin{minipage}{12.0cm}
\begin{center}
\epsfxsize=11cm
\leavevmode
\hbox{ \epsffile{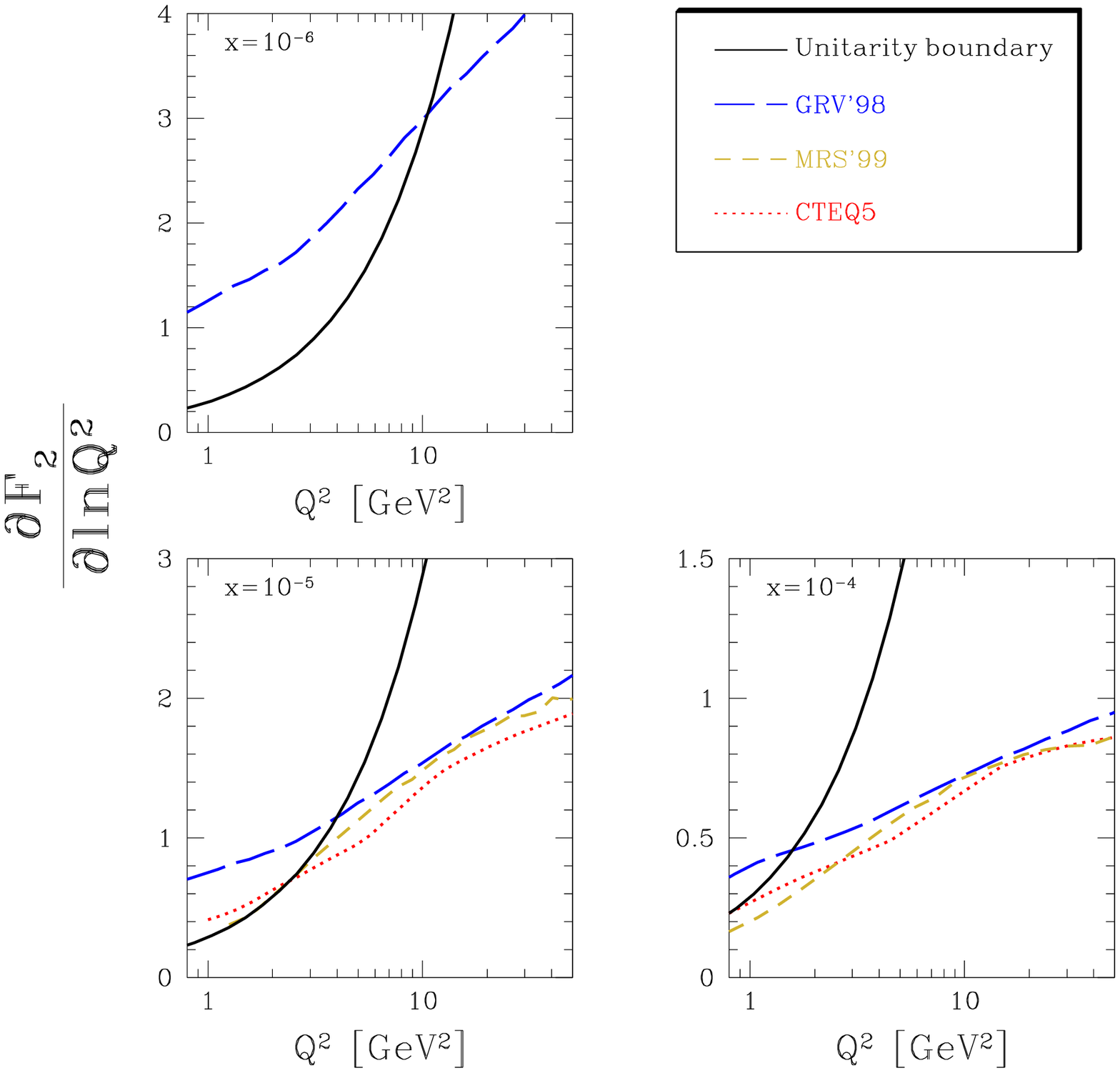}}
\end{center}
\end{minipage}
\begin{minipage}{4.0cm} 
\caption{\it The unitarity boundary and the DGLAP predictions for the
$F_2$ - slope at different values of $x$.}
\label{un}
\end{minipage}
\end{figure}
              
Fig. \ref{un}  shows that the $F_2$ slope approaches the unitarity bound at least in the GRV'98 
parameterization of the solution to the DGLAP evolution equation.
A violation of the unitarity bound in the DGLAP equation indicates  that the shadowing corrections are
 unavoidable in the THERA kinematic region. The real size of these corrections is much larger than we
 can see from the violation of the unitarity bound, since the system  starts
becoming
 dense at  densities lower than that which  follows from the unitarity
constraints. In other words, shadowing corrections lead to a  considerable
suppression in the deep inelastic structure function, at  densities lower 
than originates from the unitarity constraints.

\subsection{The unitarity bound for DIS with nuclei.}
\begin{figure}
\begin{minipage}{12.0cm}
\begin{center}
\epsfxsize=11cm
\leavevmode
\hbox{ \epsffile{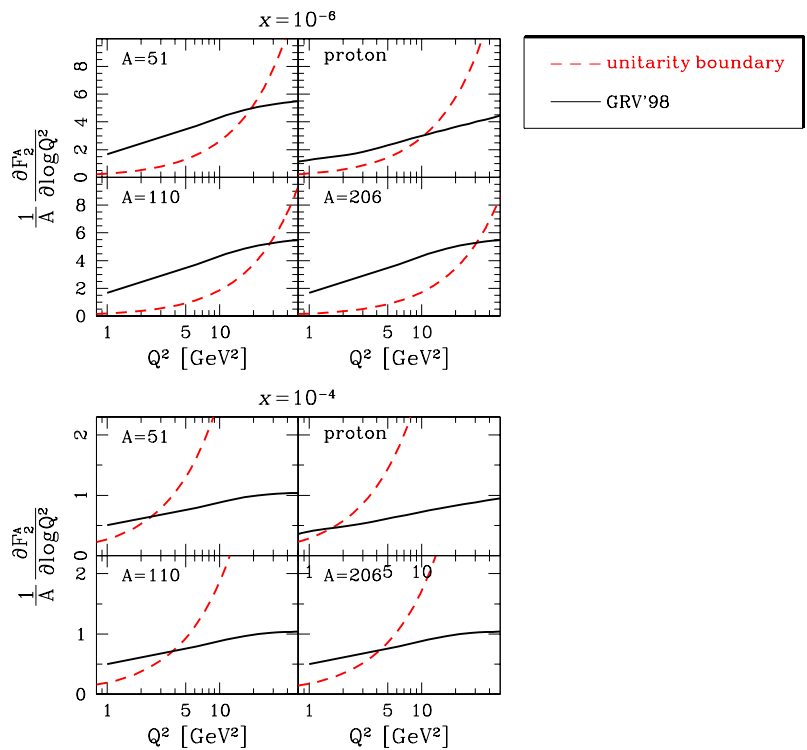}}
\end{center}
\end{minipage}
\begin{minipage}{4.0cm} 
\caption{\it The unitarity bound and the DGLAP predictions for the
$F_{2A}$ - slope at different values of $x$ and atomic number $A$.}
\label{una}
\end{minipage}
\end{figure}

\eq{UN1} looks better  for DIS with nuclei, since the   radius of the nucleus is
large  and a shrinkage
 of the diffraction peak induced by SC will be very small and we can neglect it. 
We plot the $F_{2A} = A F^{DGLAP}_{2N}$ and unitarity bound in Fig. \ref{una}.

One can see that for DIS with nuclei we should see the collective phenomena at $x \approx 10^{-4}$ and at rather
 large value of $Q^2 \approx 3  \div 5\,GeV^2$.

\subsection{Scaling violation in the $F_2$ - slope}

The careful analysis of the HERA data on the $F_2$- slope ($ \frac{\partial F_2(x,Q^2)}{\partial \ln \,Q^2}$),
given in Ref.\cite{GLMSLP}, shows that (i) our saturation model as well as
other models  of this type ( see \cite{GW}
for example ) , are  able to describe all experimental data ; and (ii) such a description cannot be very 
conclusive since other approaches are equally successful. The saturation
hypothesis leads  to 
$  \frac{\partial F_2(x,Q^2)}{\partial \ln \,Q^2}\,\,\propto\,\,Q^2\,R^2$ for $Q^2 \,\leq Q^2_s(x)$.
The HERA data \cite{H1SLP,ZEUSSLP} show such  behaviour, but we cannot
distingish this  saturation behaviour from
 the  vanishing of $F_2$ on the soft scale, which follows from the fact that
the total photoproduction cross section is finite at $ Q^2 \rightarrow 0$.
There are two reasons for this  uncertainty:
(i) the soft
 scale is not so soft  and typical transverse momentum in the  soft Pomeron could be as large as $2\, GeV$
 \cite{KL}; and (ii) the saturation scale is rather small $Q^2_s(x) = 1 \div 2 \,GeV^2$ in HERA kinematic region.
\begin{figure}
\begin{center}
\epsfxsize=14cm
\leavevmode
\hbox{\epsffile{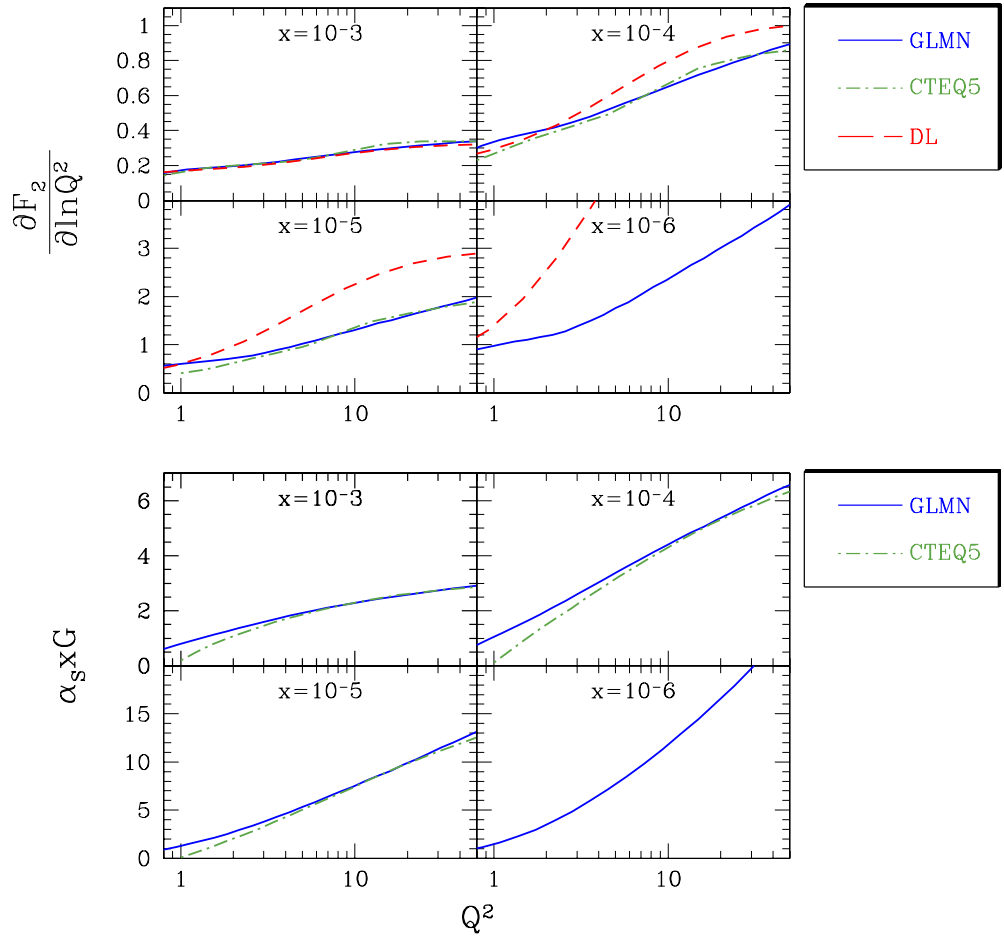}}
\end{center}
\caption{\it Predictions for the different parameterizations at fixed low values of $x$.
}
\label{slpf2}
\end{figure}

One can see from Fig. \ref{slpf2}  that THERA will allow us to distinguish
between a mixture of soft and hard Pomerons
 (DL curve in  Fig. \ref{slpf2} ) and our model for gluon saturation ( GLMN curve in  Fig. \ref{slpf2} ).
The difference between the DGLAP approach ( CTEQ5 curve in  Fig. \ref{slpf2}
) and our predictions is concentrated
 in the region of small $Q^2 \approx 1 \div 2 \,GeV^2$, but we  recall
that the corrections to  the 
 CTEQ5
 parametrization due to  high parton density 
 effects reaches the value of about 30-40\% at THERA energies. These
 estimates are an alternative way 
of saying,
that at THERA energies we expect large SC theoretically, and 
 a DGLAP
 approach can absorb these corrections in  the initial nonperturbative gluon
distributions at HERA energies, 
 but it
 would be a more difficult task in THERA kinematic region.
   
\subsection{Energy dependence of J/$\mathbf{\Psi}$ production.}
\begin{figure}
\begin{minipage}{12.0cm}
\begin{center}
\epsfxsize=11cm
\leavevmode
\hbox{\epsffile{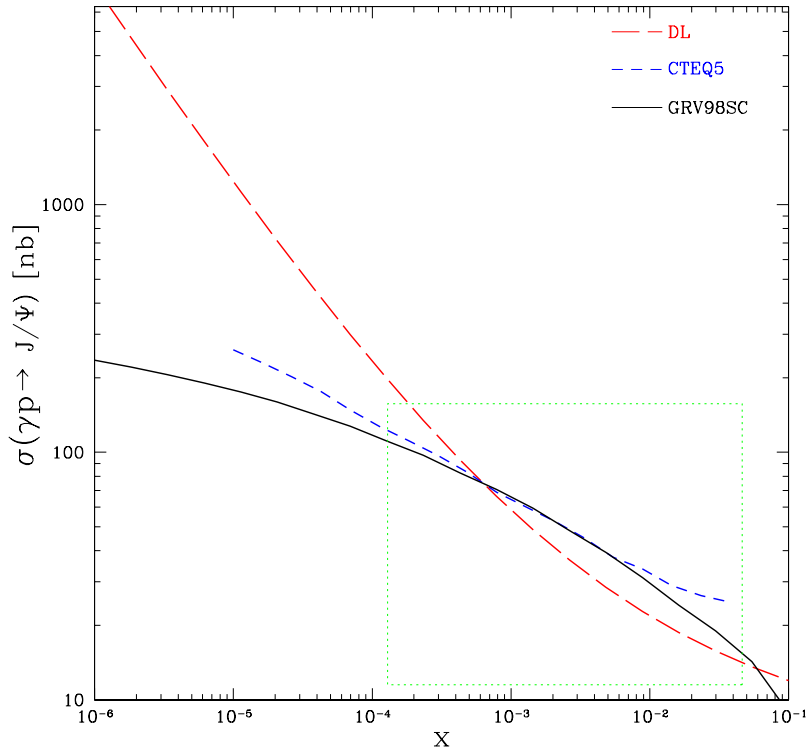}} 
\end{center}
\end{minipage}
\begin{minipage}{4.0cm}
\caption{\it A comparison between the energy dependence of different models (see text ) for the integrated cross 
section of J/$\Psi$ photoproduction. The available experimental data points are confined within the inner window.}
\label{psi}
\end{minipage}
\end{figure}

It was shown in \cite{RML, OSAKA,CALDPSI} that the energy behaviour of the J/$\Psi$ production is very sensitive
 to the value of the shadowing corrections. It turns out that the large uncertainties due to our poor knowledge of 
the wave function of vector mesons  contribute
 mostly in the normalization of the cross section,  while the energy 
slope is still a source of the information on  SC. In
Refs. \cite{OSAKA,CALDPSI} we showed that the 
shadowing corrections provide  a natural explanation of the experimental energy behaviour for 
J/$\Psi$ photo and deep inelastic  production. However, the available data do not enable us to 
exclude  explanations based on the mixture of soft and hard Pomerons, and
the idea that this process is 
  hard . Fig. \ref{psi} gives our prediction for THERA kinematic region for three approaches:
the SC calculations in our model (GRV98SC), the soft + hard Pomeron model
(DL2P)\cite{DL2P}
 and the DGLAP approach based on
 the CTEQ parametrization (CTEQ5NSC).  
                
\subsection{Shrinkage of diffraction peak for production of vector mesons.}

The experimentally observed shrinkage of the diffraction peak in photoproduction of J/$\Psi$\cite{PSIDATA}
is a direct indication that this process is not a simple hard process that can be described in the DGLAP approach
. Indeed,  one of the most well established properties of the hard processes is the fact that the $t$-dependence 
is independent of energy ($x$) (see for example Ref. \cite{GLMVP}. There are two possible explanations: (i) the
 first one is the SC which lead to $x$-dependence of the $t$-slope\cite{GLMVP}
and (ii) the second is  based on the contamination of the J/$\Psi$ production by the soft processes for
 which the shrinkage of the diffraction peak is a phenomenon that  is  well established
both theoretically and experimentally.

It should be stressed that the above  two approaches have different
predictions for  the  $x$-dependence:
SC lead to the $t$-slope which increases at higher energies ( lower $x$), since the contribution of SC grows with
 energy. For the mixture of soft and hard processes, the role of soft ones diminishes at higher energies, and 
as a result the value of effective $\alpha'_{eff}$ decreases \cite{MSF}.    

\begin{figure}
\begin{minipage}{12.0cm}
\begin{center}
\epsfxsize=11cm
\leavevmode
\hbox{ \epsffile{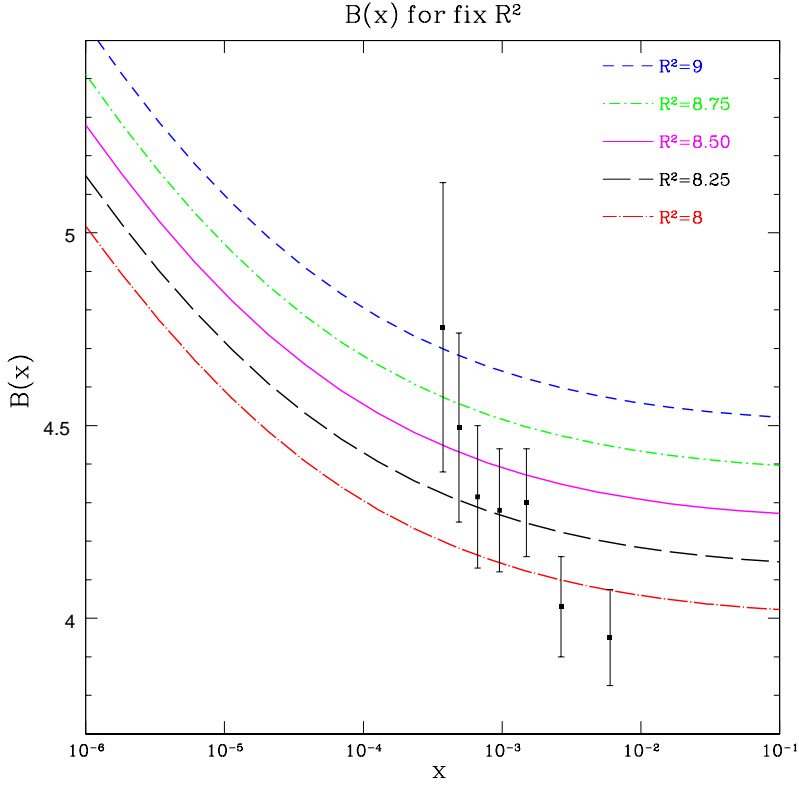}}
\end{center}  
\end{minipage}
\begin{minipage}{4.0cm} 
\caption{\it The energy ($x$) dependence of the forward differential slope of
J/$\Psi$ photoproduction.ZEUS data
 \protect\cite{PSIDATA} and our model calculation with several values of
$R^2$. From this picture we chose 
the value of $R^2 = 8.5\,GeV^{-2}$ for the typical proton size in our model.}
\label{psislp}
\end{minipage}
\end{figure}

In Fig.~\ref{psislp} one can see this effect for our model. This figure also shows
 that we cannot describe the ZEUS 
data regarding  the value of the $t$-slope. The reason for this may be
 due to our under estimating the value of SC in our model. However, one
lesson we can learn from Fig.~\ref{psislp}: 
THERA will clarify the question which  mechanism works. The important thing
to emphasize once more is that the measurement of the shrinkage  of the
$t$-slope will provide reliable
 information on the deviation from the simple DGLAP approach.It is  especially important to observe 
such  shrinkage in the DIS diffraction production of  J/$\Psi$ and other vector mesons.

\subsection{Maxima in ratios}
\begin{figure}
\begin{tabular}{l l }
\psfig{file=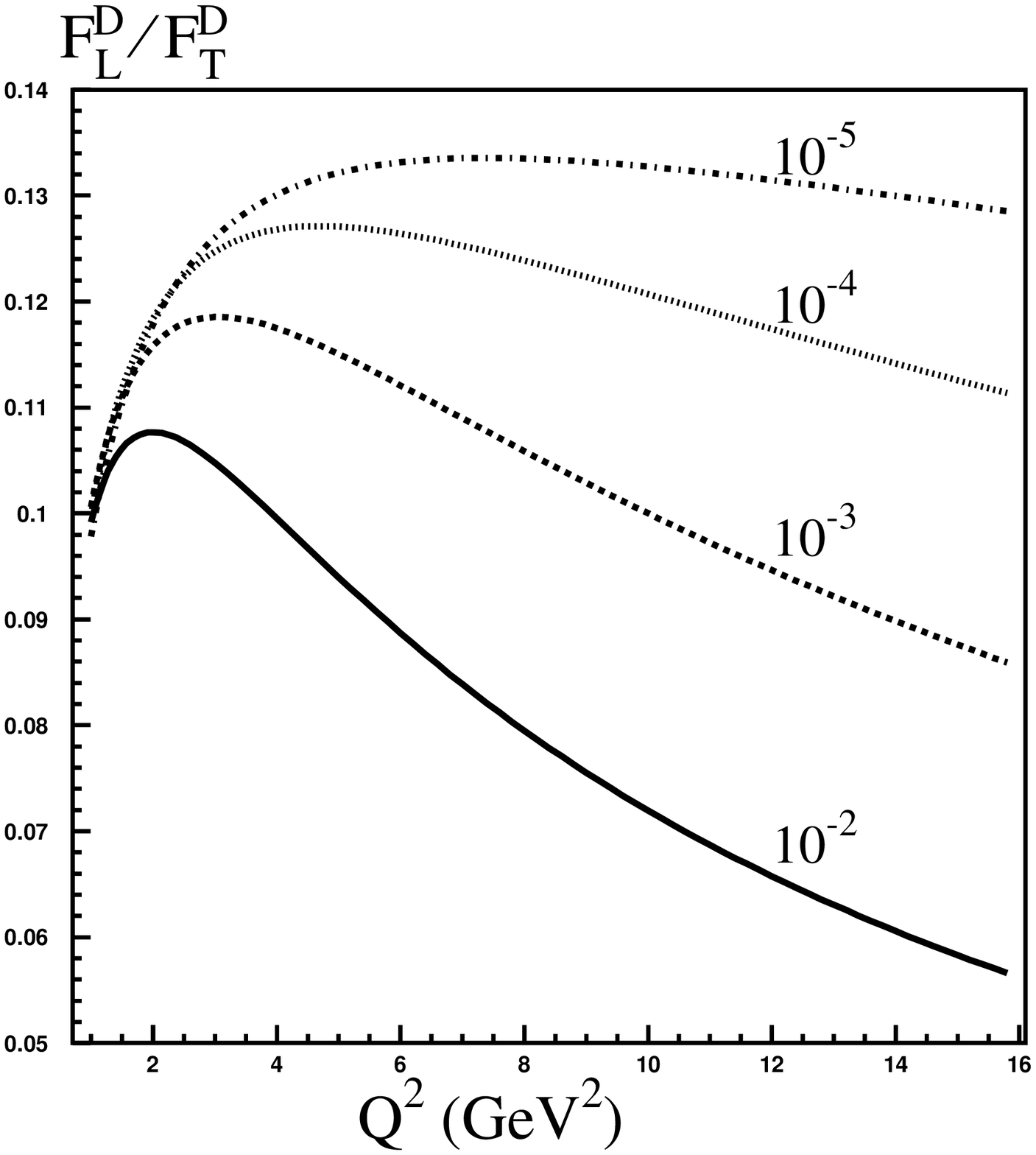,width=8.5cm} &
\psfig{file=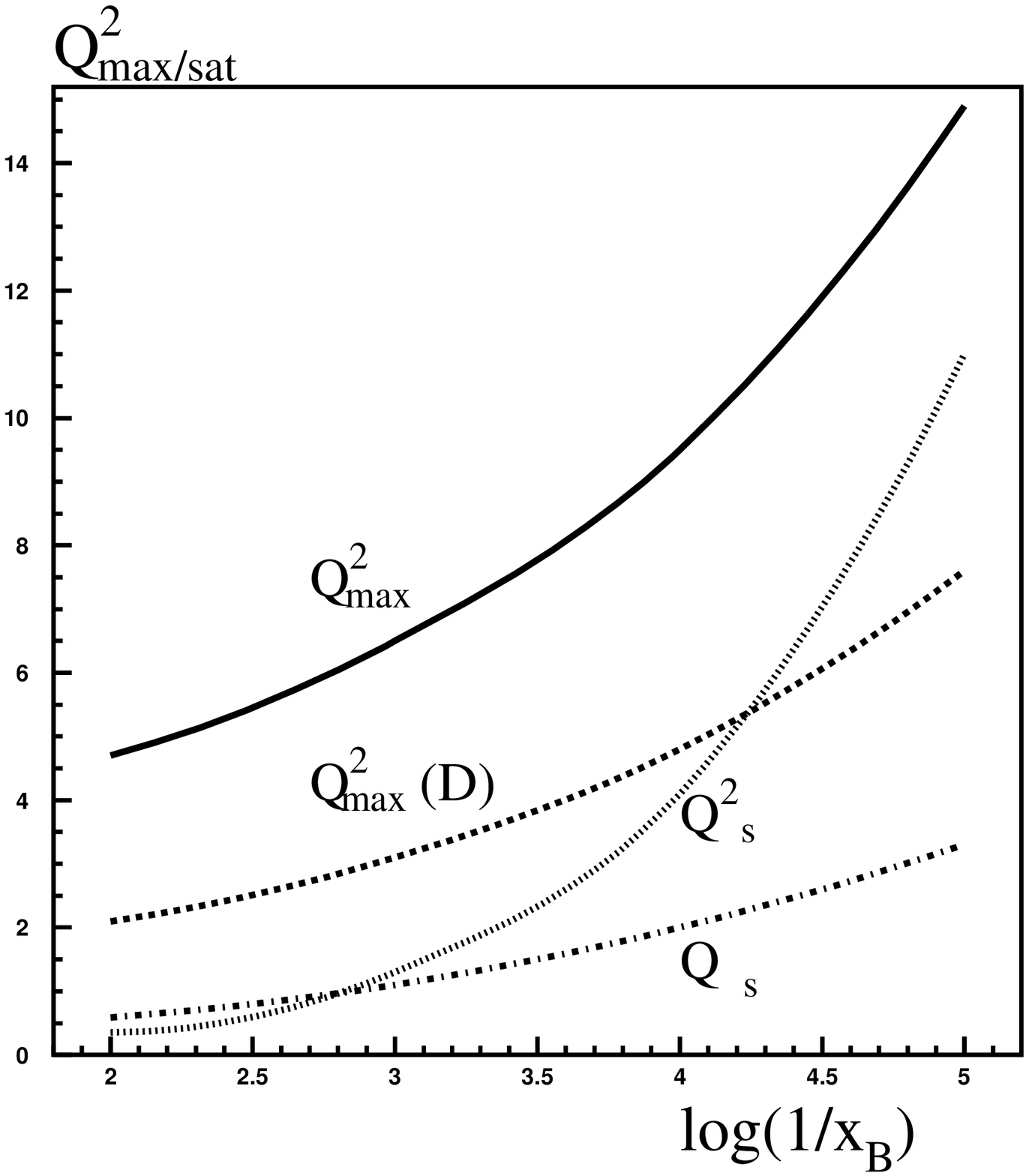,width=8.5cm}\\
\end{tabular}
\caption{\it The ratio of $F^D_L/F^D_T$ as function of $Q^2$ at different
values of $x$ and the behavior of $Q^2_{max}(x)$ as function of $x$.}
\label{max}
 \end{figure}

 Preparing this paper we tried to find  improved observables which will be sensitive to the saturation scale.
We study the $Q^2$ behaviour of the ratios:$F_L/F_T$ and $F^D_L/F^D_T$ for longitudinal and transverse structure
 function for inclusive DIS and for diffraction in DIS \cite{GLMHT}. In Fig.~\ref{max} some examples of these
 ratios are plotted. We found that these ratios have maxima at $Q^2 = Q^2_{max}(x)$ which increases with $x$ as
 $x \rightarrow 0$. It appears that $Q_{max}(x)$ is a simple function of the
saturation scale $Q_s(x)$.
In the  THERA kinematic region $Q^2_{max}(x)$ is large $Q^2_{max}(x) \approx 6
\div 7 \,GeV^2$. Such a large value
 of $Q^2_{max}(x)$ makes our calculation more reliable, and we expect
that the measurement of this maxima in the THERA kinematic region will enable
us to extract the value of the saturation scale from the experimental data. 

\subsection{Higher twist contribution.}

\begin{figure}
%\begin{flushleft}
\begin{tabular}{c c c}
\multicolumn{3}{c}{\rule[-3mm]{0mm}{4mm} $ F_L(Q^2)$}\\ 
$x_B=10^{-3}$ & $x_B=5\cdot 10^{-4}$ & $x_B=10^{-5}$\\
\psfig{file=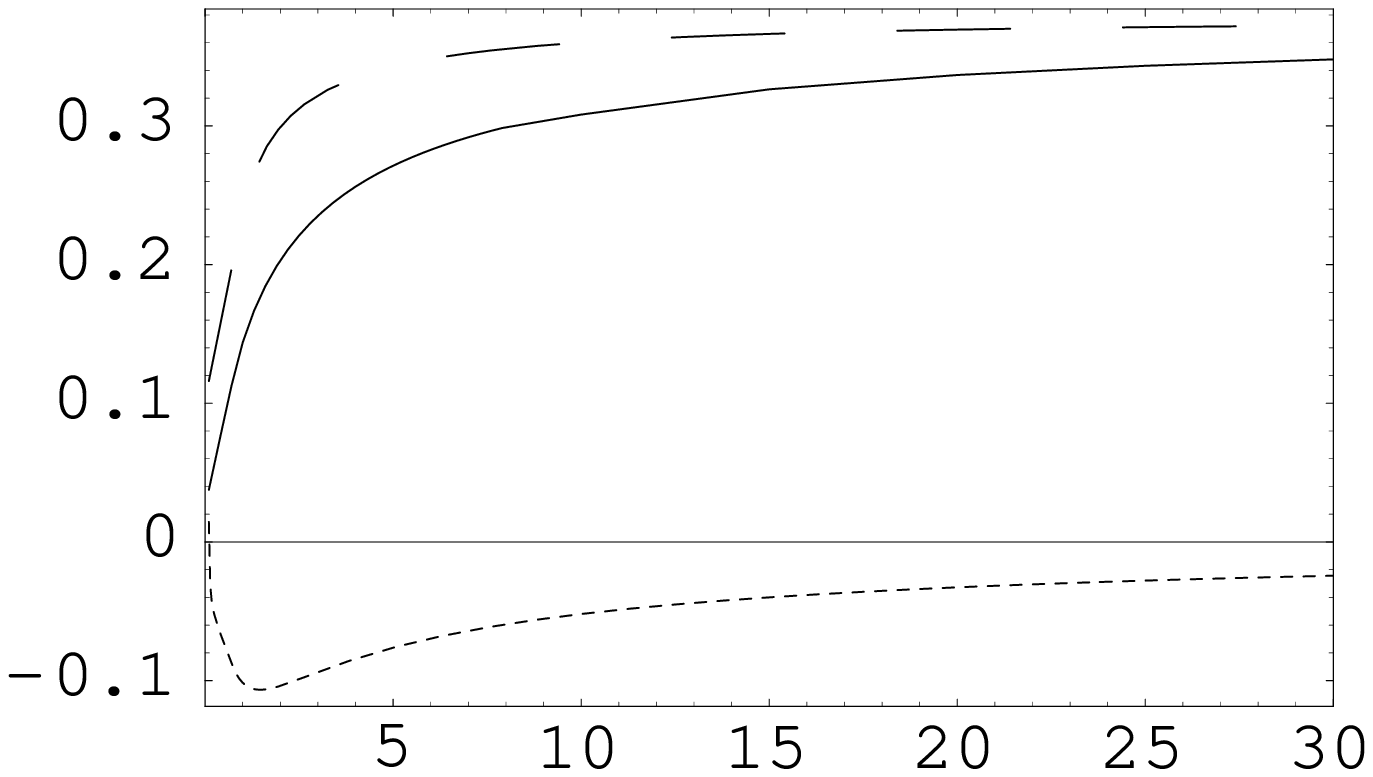,width=50mm,height=45mm} &
 \psfig{file=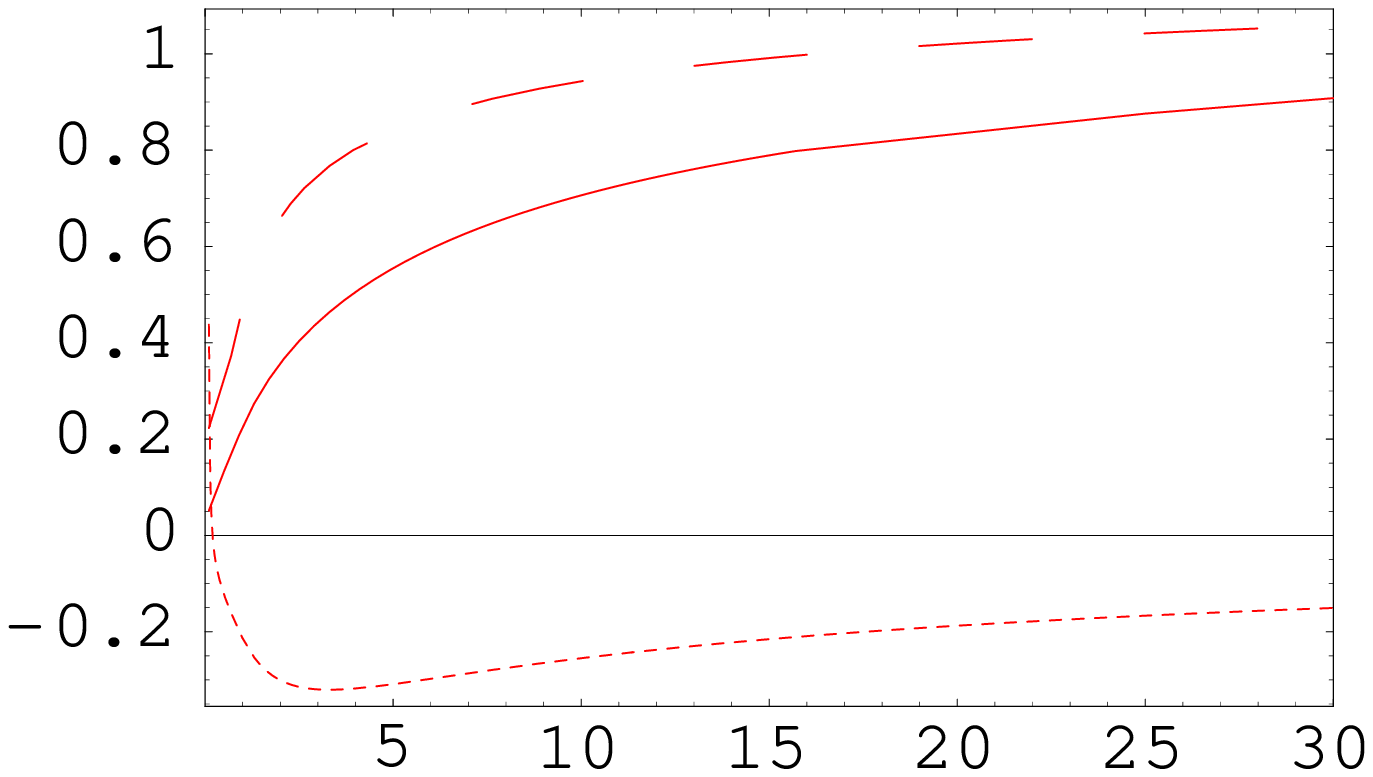,width=50mm,height=45mm} &
\psfig{file=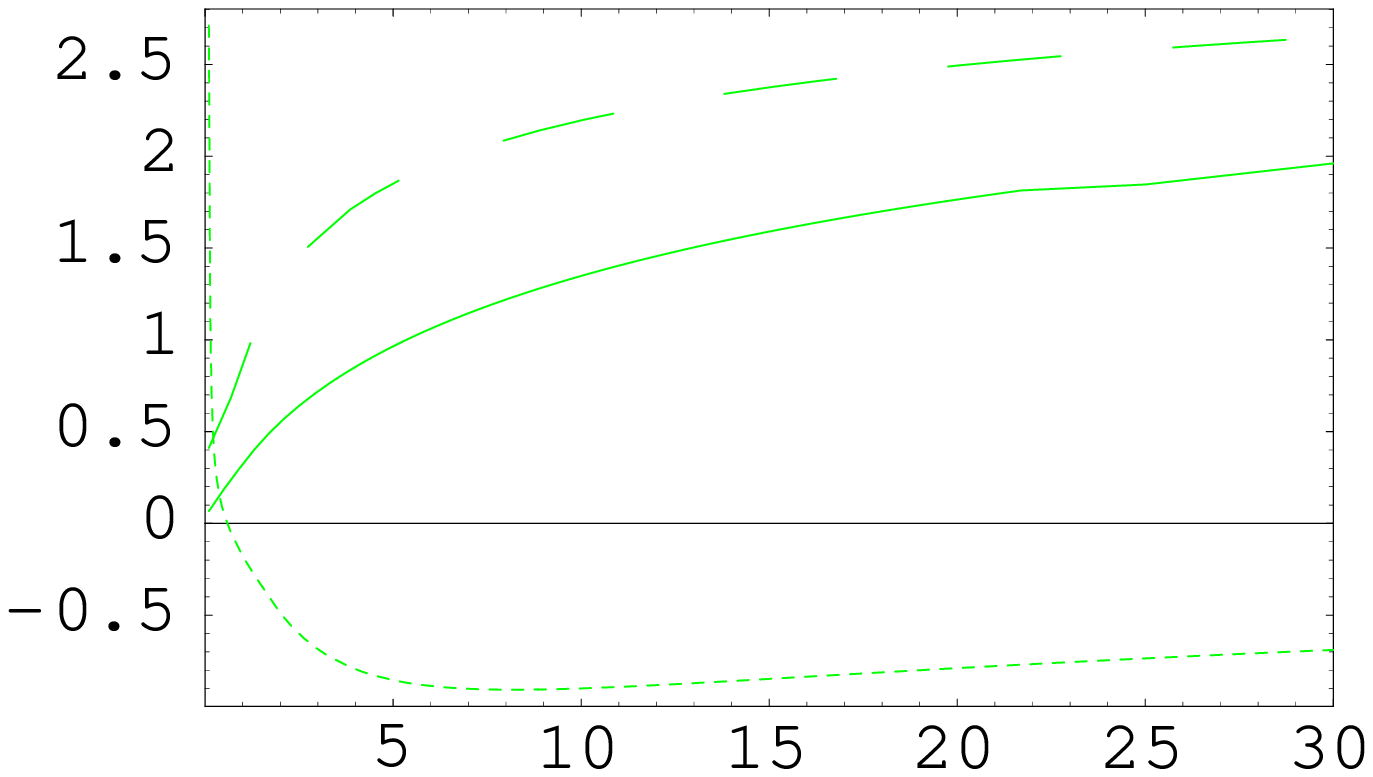,width=50mm,height=45mm}\\ 
\multicolumn{3}{c}{\rule[-3mm]{0mm}{4mm} $ F_T(Q^2)$}\\ 
$x_B=10^{-3}$ & $x_B=5\cdot 10^{-4}$ & $x_B=10^{-5}$\\
\psfig{file=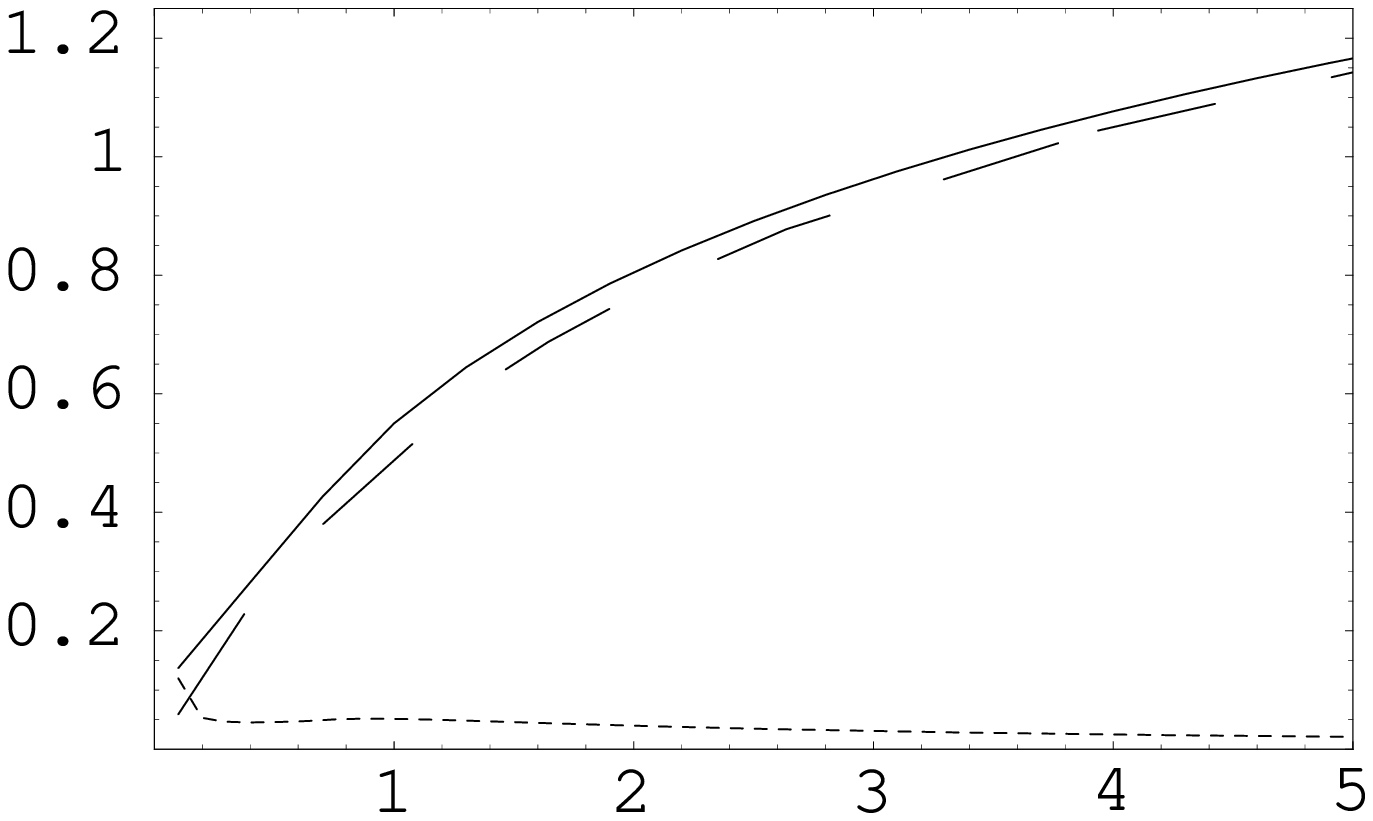,width=50mm,height=45mm} &
\psfig{file=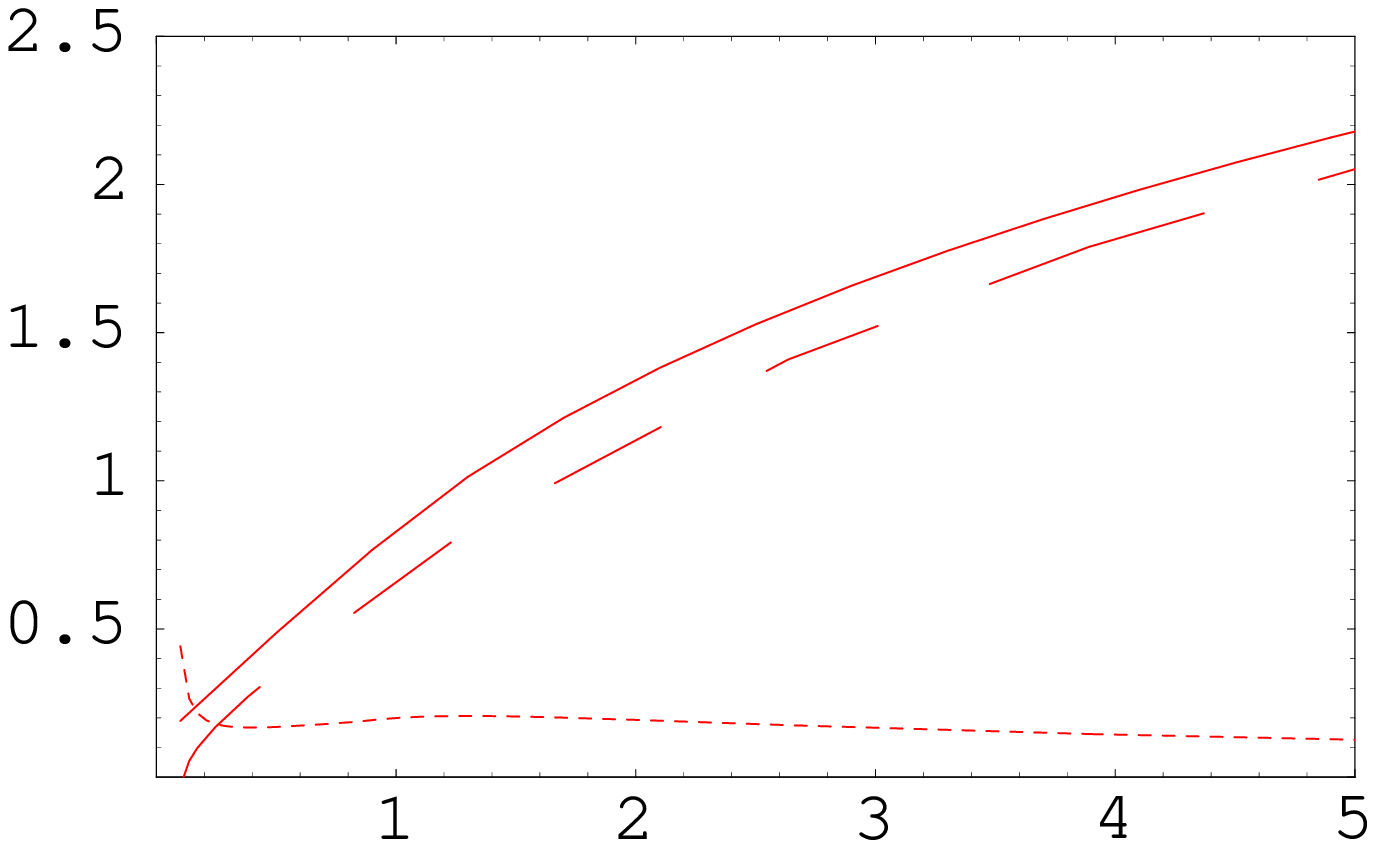,width=50mm,height=45mm} &
\psfig{file=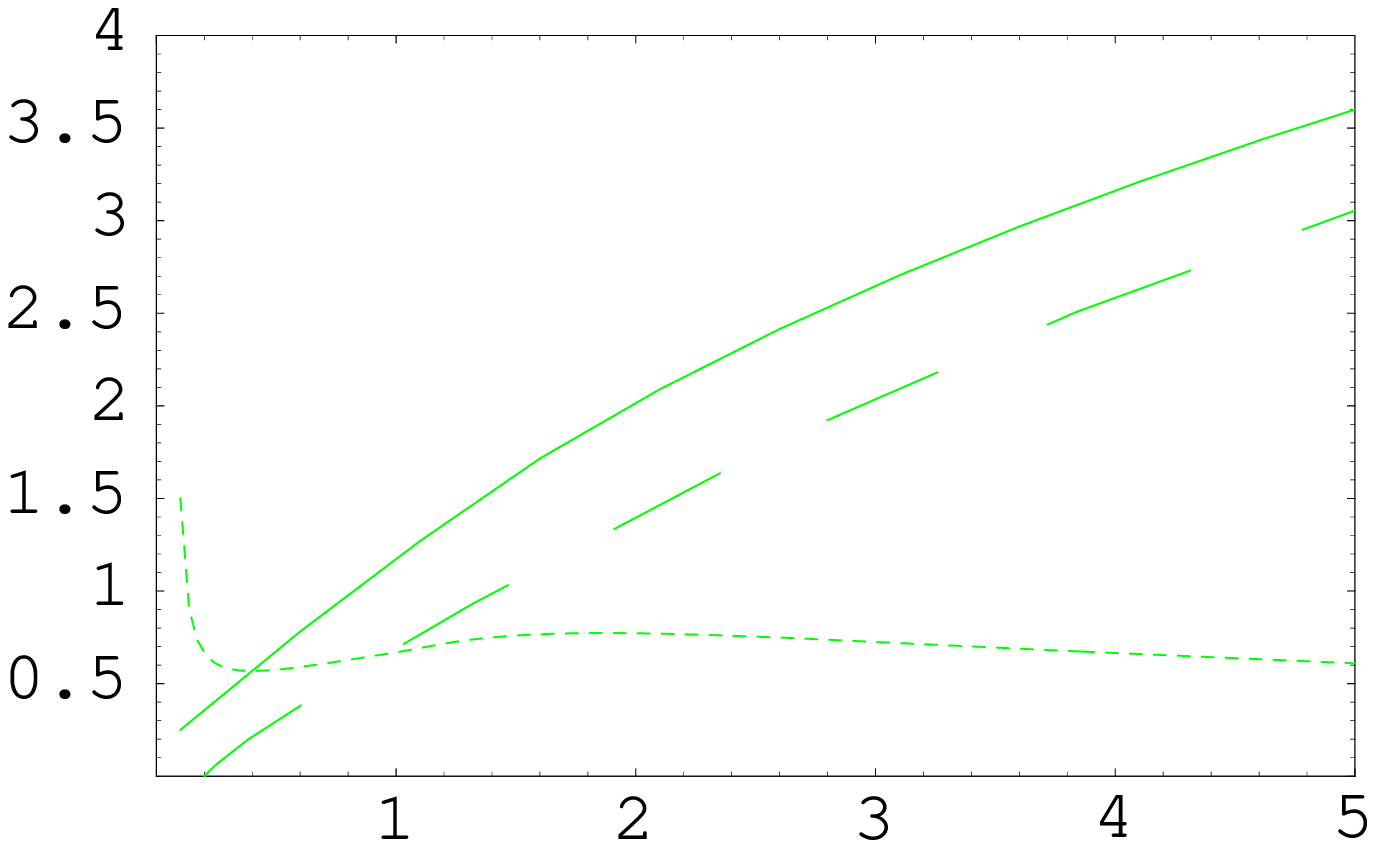,width=50mm,height=45mm}\\
 \multicolumn{3}{c}{\rule[-3mm]{0mm}{4mm} $ F^D_L(Q^2)$}\\ 
$x_B=10^{-3}$ & $x_B=5\cdot 10^{-4}$ & $x_B=10^{-5}$\\
\psfig{file=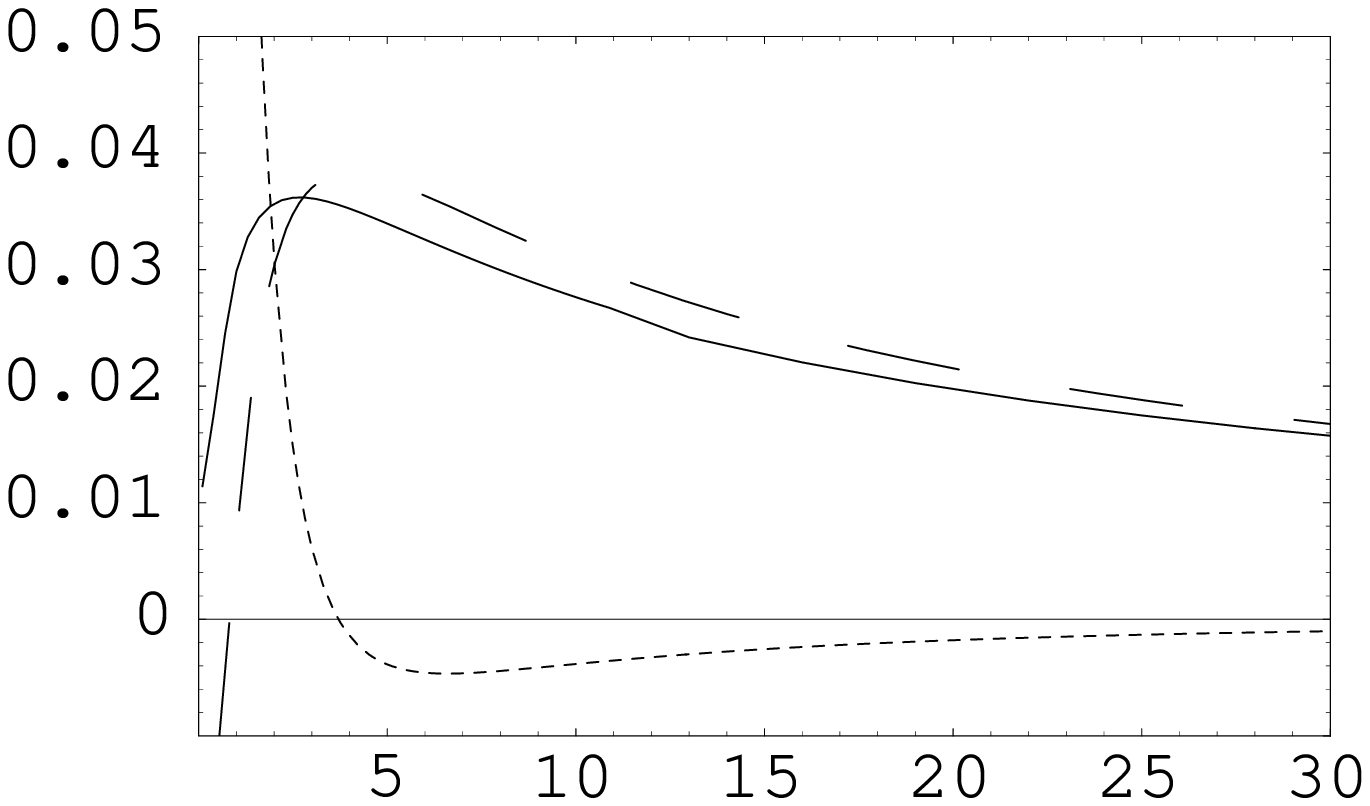,width=50mm,height=45mm} &
\psfig{file=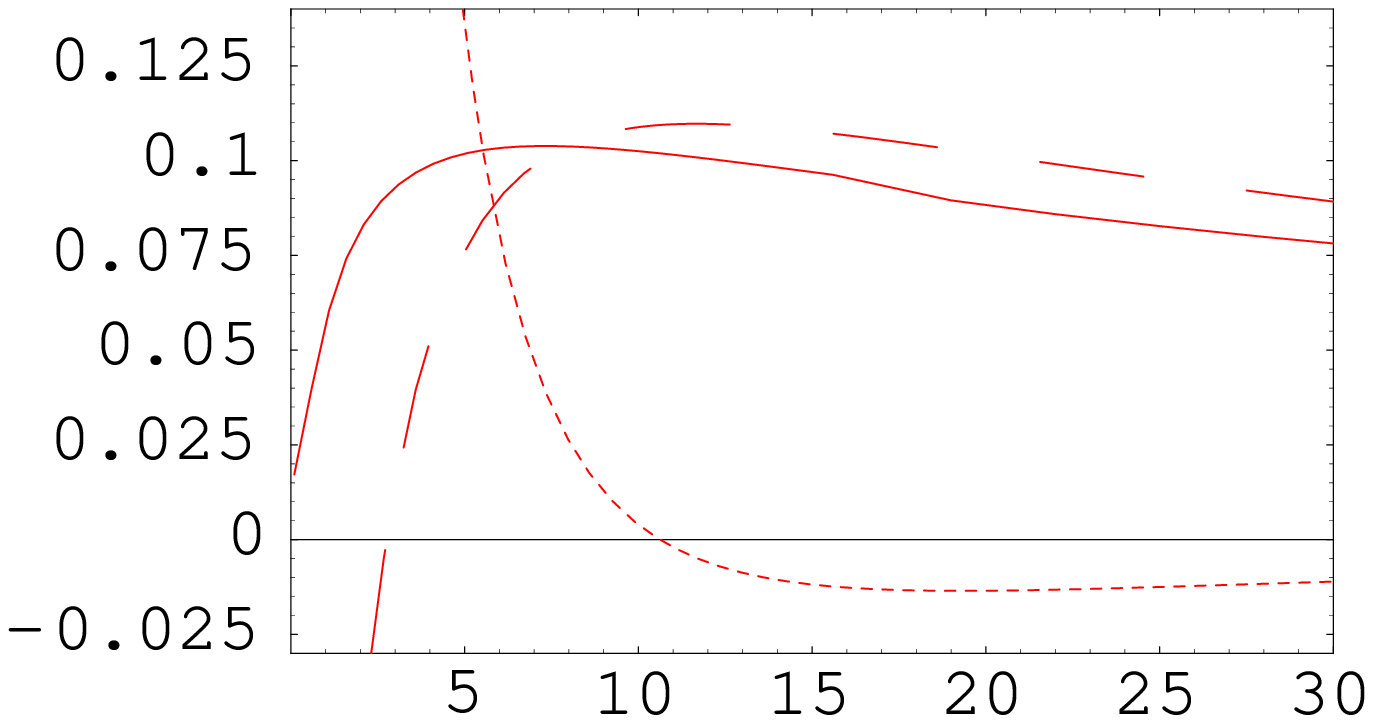,width=50mm,height=45mm} &
\psfig{file=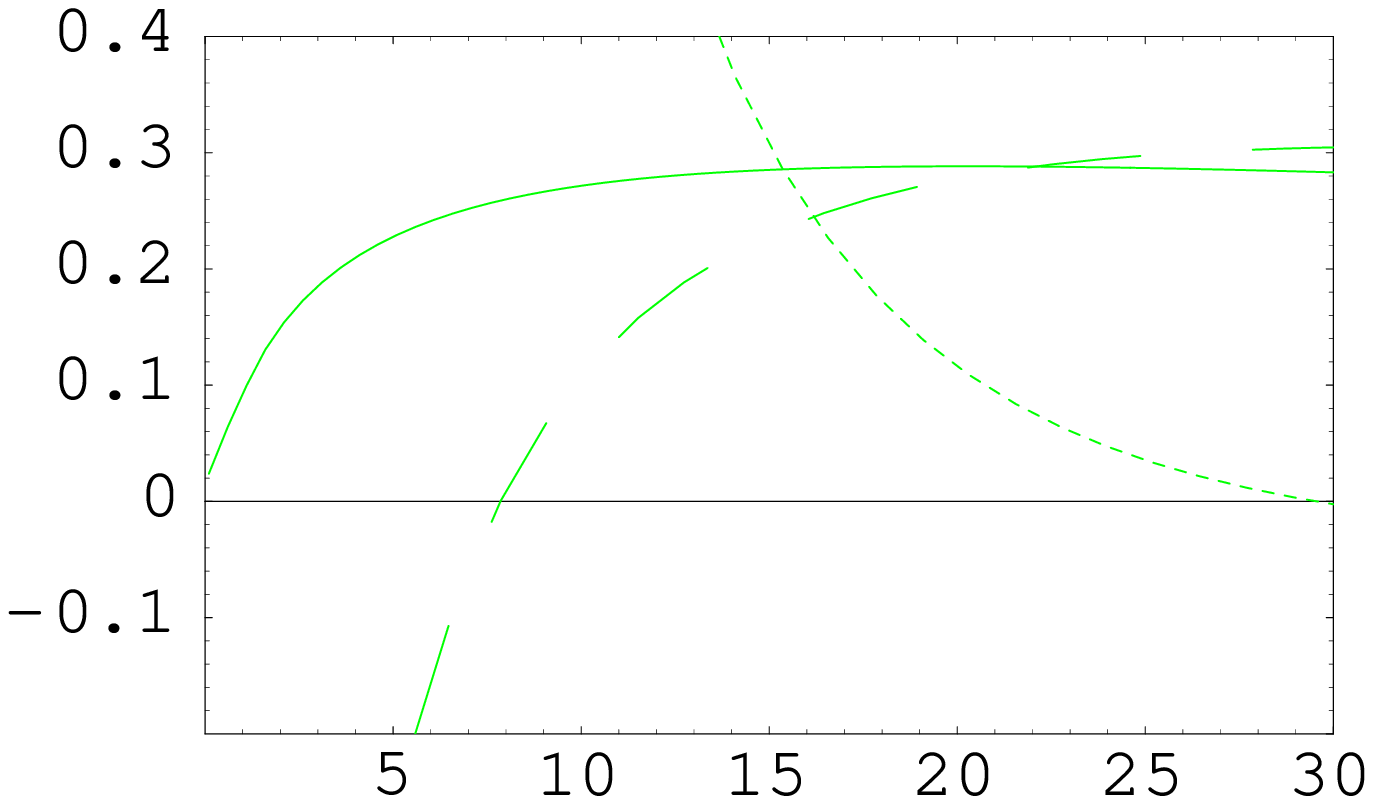,width=50mm,height=45mm}\\  
\multicolumn{3}{c}{\rule[-3mm]{0mm}{4mm} $ F^D_T(Q^2)$}\\ 
$x_B=10^{-3}$ & $x_B=5\cdot 10^{-4}$ & $x_B=10^{-5}$\\
\psfig{file=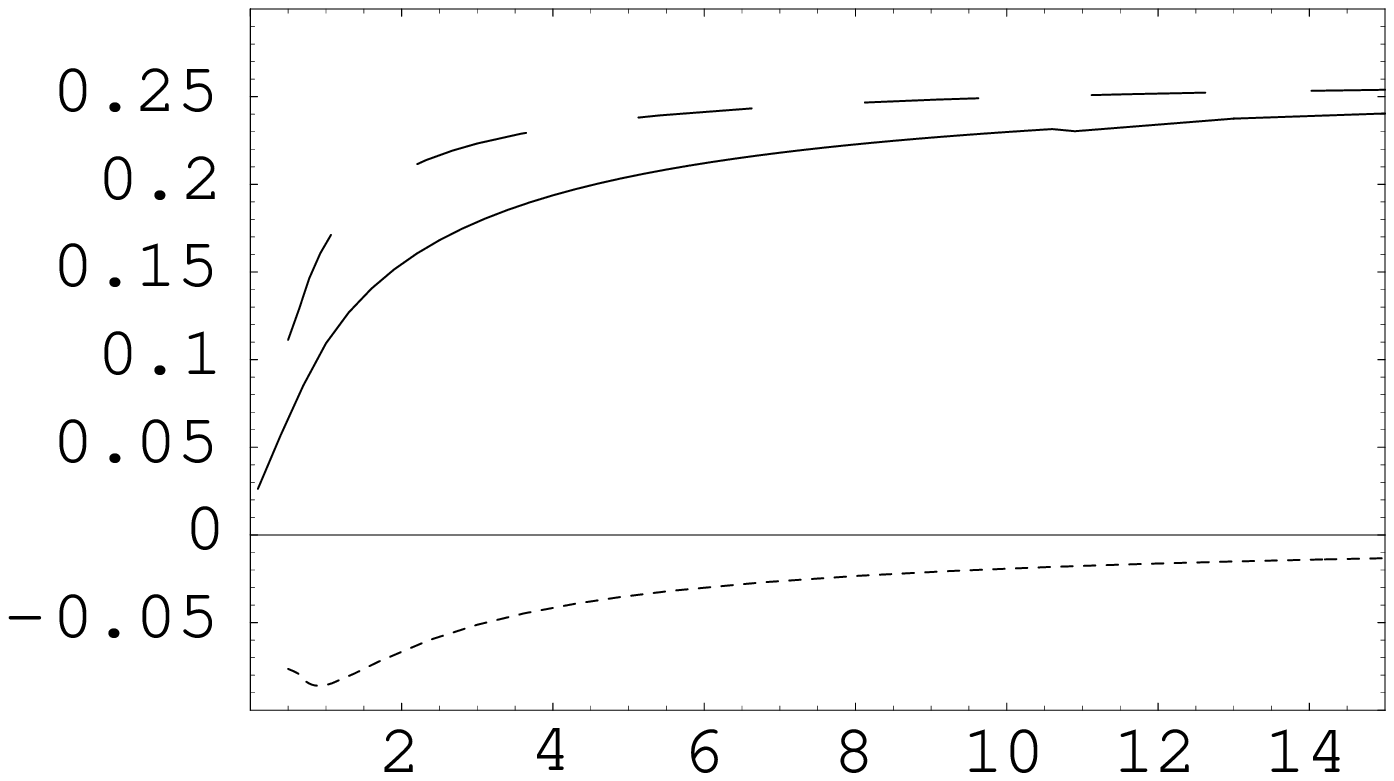,width=50mm,height=45mm} &
\psfig{file=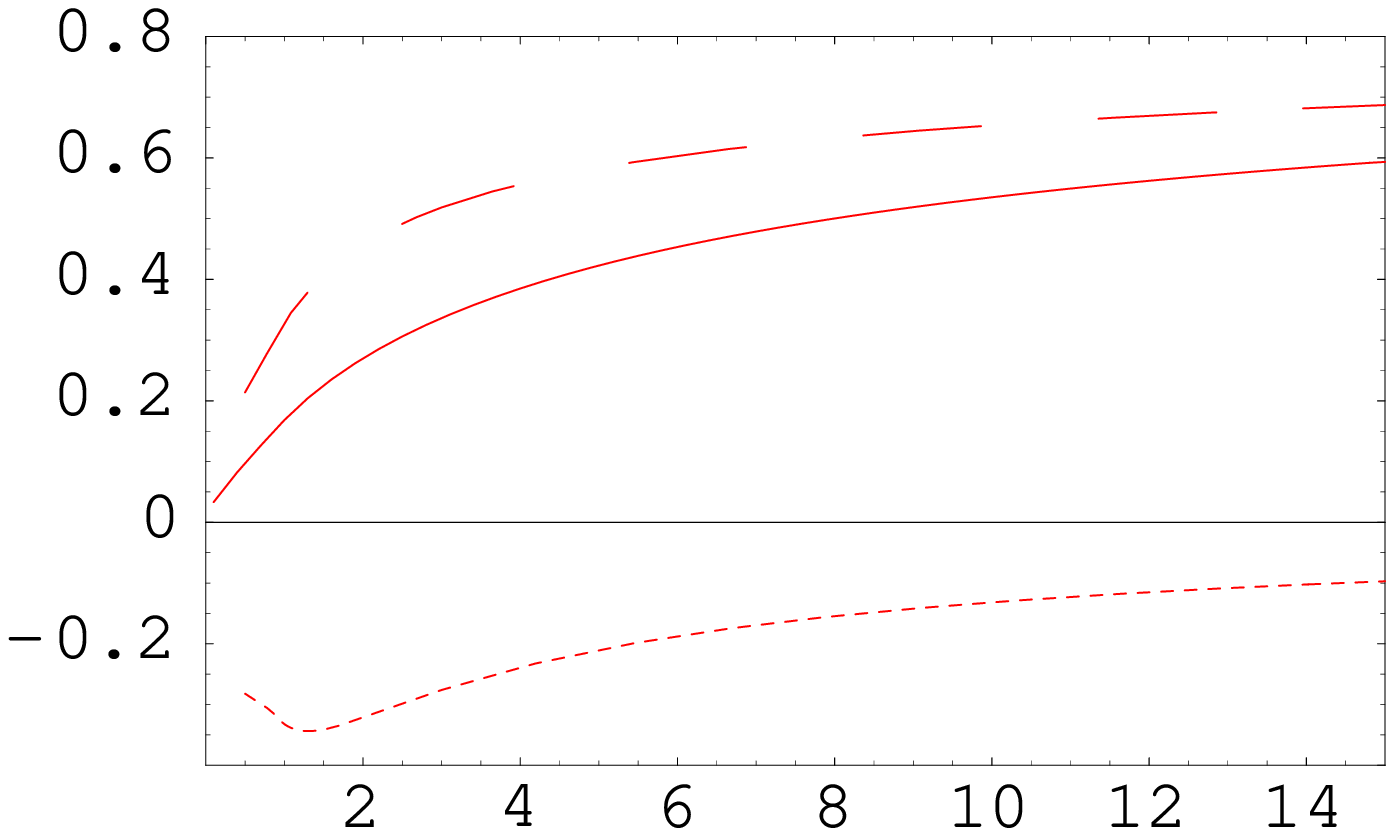,width=50mm,height=45mm} &
\psfig{file=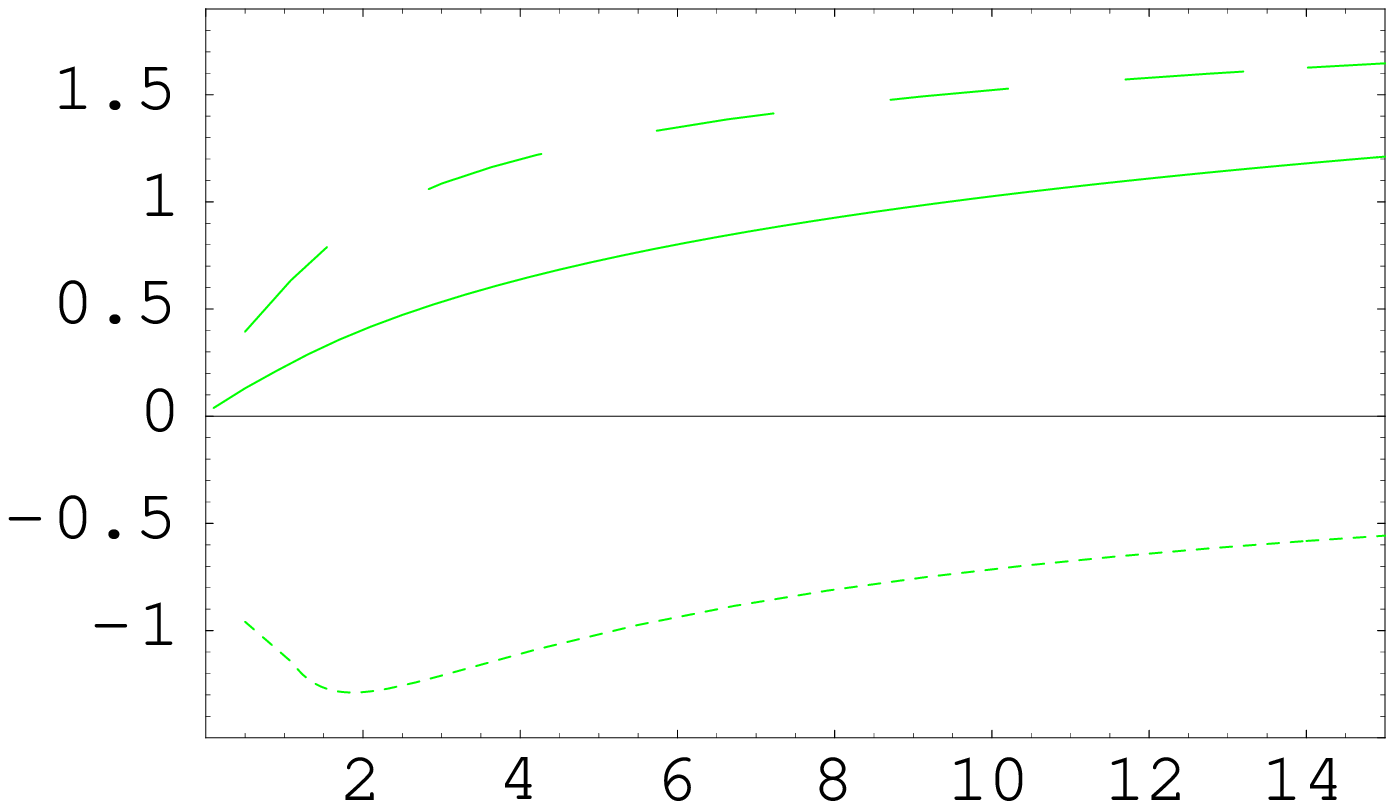,width=50mm,height=45mm}\\
\end{tabular}
%\end{flushleft}
\vspace{-1cm}   
 \caption{\it Different twist contributions
to the various structure functions for DIS on the proton:
leading twist (at high $Q^2$) -- dashed line,
next-to-leading -- dotted one, exact structure function --
solid curve.}
\label{htw}
\end{figure}

One of the most challenging problem of QCD is to understand the higher twist contributions. The present approach
 to DIS is based on two main ideas: (i) the DGLAP evolution equation for leading twist contributions and (ii) the
 firm belief that higher twist contributions are small in the whole kinematic
region, when  we start  QCD evolution 
from the large value of $Q^2 = Q^2_0 \approx 1 - 4\, GeV^2$. In recent years it has been proven that there is no
 ground  for such an assumption. It was found \cite{HT} that the anomalous dimension for the higher twists is much
 larger than 
for the leading one in the region of low $x$. It turns out that if we write the deep inelastic structure function
 in the form
$$
F_2(x,Q^2) \,\,=\,\,F^{LT}_2(x,Q^2)\,\,+\,\,\frac{M^2}{Q^2} \,F^{HT}_2(x,Q^2)$$
$F^{HT}_2(x,Q^2)\,\,\propto\,\,F^{LT}_2(x,Q^2) \,\times\,xG(x,Q^2)$ at $x \,\rightarrow\,0$. Therefore, the
 experimental observation of the higher twist contribution, is one of the most challenging and important problems
 in DIS, as well as in QCD at large. The attractive feature of our model is
the fact that it leads to  higher
 twist contributions in accord with   known theoretical information. Our calculations confirm the result of
 Ref. \cite{HTM} that there is almost a full cancellation of the higher twist
contributions
 in $F_2$ in spite of the fact that they give substantial contributions
separately to $F_L$ and $F_T$ as well as to
 $F^D$. In the  THERA kinematic region (
 at $x \approx 10^{-5}$)  we expect  the higher twist contributions to be of the same order
as the leading twist  at sufficiently high value of $Q^2$ (see
Fig.~\ref{htw})  This high value of $Q^2$ insures
 us that our calculations are reliable. Therefore, we believe that THERA has a good chance to measure higher
 twist terms  and a new era of DIS will open, that will include a
systematic study of higher twist contributions.
\section{Resume}
We presented here our estimates for  the possible manifestation of
saturation in the  THERA kinematic region.
We believe that HERA has reached a new QCD regime: the high parton density QCD domain \cite{AMIRIM},
 where  incorporating new collective phenomena
is   essential for understanding it's  physics. We argue here that data from THERA will
be able to show  that we have  reached this
 new regime, and will allow a systematic study of the QCD parton system with large parton density. 

We hope that our estimates will help to plan  the experimental strategy
for   the  THERA project.

\section*{Acknowledgements}

The authors are very much indebted to our coauthors and friends with
whom we discussed our approach on  a everyday basis Ian Balitsky, Jochen
Bartels ,
Krystoff Golec Biernat, Larry
McLerran, Dima Kharzeev, Yuri Kovchegov and  Al Mueller for their help and
fruitful discussions on the subject. E.G. ,  E. L. U.M. and K.T.  thank BNL
Nuclear
Theory Group and  DESY Theory group
for their hospitality and
creative atmosphere during several stages of this work.
     
 This research was supported in part by the BSF grant $\#$
9800276, by the GIF grant $\#$ I-620-22.14/1999 
  and by
Israeli Science Foundation, founded by the Israeli Academy of Science
and Humanities.

\end{document}